\documentclass[12pt,preprint]{aastex}

\def\bc{\begin{center}}
\def\ec{\end{center}}

\shorttitle{Internal Properties of UCDs in the Virgo Cluster}
\shortauthors{Evstigneeva et al.}

\received{2006 August 9}
\begin{document}


\title{Internal Properties of Ultra-Compact Dwarf Galaxies \\
    in the Virgo Cluster}


\author{E.~A.~Evstigneeva}
\affil{Department of Physics, University of Queensland, QLD 4072, Australia}
\email{katya@physics.uq.edu.au}

\author{M.~D.~Gregg\altaffilmark{1}}
\affil{Department of Physics, University of California, Davis, CA 95616, USA}
\email{gregg@igpp.ucllnl.org}

\author{M.~J.~Drinkwater}
\affil{Department of Physics, University of Queensland, QLD 4072, Australia}
\email{m.drinkwater@uq.edu.au}

\and

\author{M.~Hilker\altaffilmark{2}}
\affil{Argelander-Institut f\"ur Astronomie, Universit\"at Bonn, Auf dem H\"ugel 71, 53121 Bonn, Germany}
\email{mhilker@eso.org}


\altaffiltext{1}{Institute for Geophysics and Planetary Physics, Lawrence Livermore National Laboratory, L-413, Livermore, CA 94550, USA}
\altaffiltext{2}{European Southern Observatory, Karl-Schwarzschild-Str. 2, 85748 Garching bei M\"unchen, Germany} 


\begin{abstract}
We present new imaging and spectroscopic observations of
  six ultra-compact dwarf (UCD) galaxies in the Virgo Cluster, along with
  re-analysed data for five Fornax Cluster UCDs. These are the most
  luminous UCDs: $-14<M_V<-12$ mag. Our {\it Hubble Space Telescope}
  imaging shows that most of the UCDs have shallow or steep cusps in
  their cores; only one UCD has a flat ``King'' core.  None of the
  UCDs show tidal cutoffs down to our limiting surface
  brightness. Spectroscopic analysis shows that Virgo UCDs are old
  (older than 8 Gyr) and have metallicities in the range from 
  [Z/H] = -1.35 to +0.35 dex. Five Virgo UCDs have super-solar
  [$\alpha$/Fe] abundance ratios and one Virgo UCD has a solar
  abundance ratio.  The super-solar [$\alpha$/Fe] abundances are
  typical of old stellar populations found in globular clusters and
  elliptical galaxies.  We find that Virgo UCDs have structural and dynamical 
  properties similar to Fornax UCDs. The Virgo and Fornax UCDs all
  have masses $\approx 2-9 \times 10^7M_\odot$ and mass-to-light
  ratios $\approx 3-5 \, M_\odot/L_{\odot,V}$.  The dynamical
  mass-to-light ratios for Virgo UCDs are consistent with simple
  stellar population model predictions: the Virgo UCDs do not require
  dark matter to explain their mass-to-light ratios.
  We conclude that the internal properties of Virgo UCDs are consistent with
  them being the high-mass/high-luminosity extreme of known globular
  cluster populations. We refrain from any firm conclusions on  
  Fornax UCD origins until accurate age, metallicity and $\alpha$-abundance 
  estimates are obtained for them. Some of our results, notably the fundamental
  plane projections are consistent with the formation of UCDs by the
  simple removal of the halo from the nuclei of nucleated dwarf
  galaxies. However the ages, metallicities and abundances for Virgo
  UCDs are not consistent with this simple stripping model.  It might
  be consistent with more sophisticated models of the stripping
  process that include the effects of gas removal on the chemical
  evolution of the nuclei.  
\end{abstract}





\section{Introduction}

Recent spectroscopic surveys of the Fornax and Virgo galaxy clusters
have revealed a new class of compact stellar system, ``ultra-compact
dwarf galaxies'' (UCDs) with properties intermediate between the larges
globular clusters and the smallest dwarf galaxies. The defining
properties of the first UCDs found are that they are significantly
more luminous than most known globular clusters ($-14<M_V<-12$), but
they are mostly unresolved in ground-based sky survey images. In this
paper we present detailed observations of a new sample of UCDs in the
Virgo cluster in order to test hypotheses for the formation of these
objects.

The Fornax Cluster UCD objects were discovered independently in
studies of globular cluster systems around the central galaxy NGC~1399
(Minniti et al.\ 1998, Hilker et al.\ 1999) and in studies of compact
dwarf galaxies in the cluster (Drinkwater et al.\ 2000, Phillipps et
al.\ 2001). Confirmed UCDs have subsequently been found in the Virgo Cluster
(Ha\c{s}egan et al.\ 2005, Jones et al.\ 2006) and UCD candidates
identified in the more distant cluster Abell~1689 (Mieske et al.\
2004). At low luminosities ($M_V>-12$) the distinction between UCDs and
globular clusters is not clear (see discussions in Drinkwater et al.\
2004, Mieske et al.\ 2006), but in this paper we focus on the most
luminous objects ($-14<M_V<-12$).

Given their intermediate nature, most formation hypotheses for UCDs
relate them to either globular clusters or dwarf galaxies. They could
be highly luminous intra-cluster globular clusters (e.g.\ Mieske et
al.\ 2002).
Alternatively, UCDs may result from the tidal disruption of
nucleated dwarf elliptical galaxies. This process can leave the
nucleus intact on intra-cluster orbit (Bekki et al.\ 2001, 2003)
as a UCD. Other formation scenarios include UCDs being the evolved
products of massive super starclusters formed in galaxy mergers
(Fellhauer \& Kroupa 2002), or primordial objects (Phillipps et al.\
2001).

It is hard to provide definitive observational tests of these
different scenarios for UCD formation. This is partly because the
various scenarios do not always make very different predictions, but
the observational picture also remains far from complete. The
observational tests can broadly be divided into statistical population
studies (e.g.\ the distribution of UCDs compared to other objects) and
detailed studies of the internal properties (e.g.\ internal velocity
dispersion) of the UCDs. In this paper we focus on the latter.

Our previous investigations have focused on the Fornax Cluster
UCDs. We used {\em Hubble Space Telescope} (HST) imaging and ESO Very Large
Telescope spectroscopy to compare the five original UCDs to globular
clusters and nucleated dwarf galaxies (Drinkwater et al.\ 2003). We
concluded that UCDs were distinct from known globular cluster
populations as they followed a different relation between internal
velocity dispersion and luminosity. The UCD properties were, however
consistent with the threshing model in which the dwarf galaxy nuclei
survived tidal disruption with no significant change to their
luminosity or velocity dispersion.

The tidal disruption hypothesis was also supported by Ha\c{s}egan et al.\
(2005) in their analysis of 10 compact ``dwarf-globular transition
objects'' found in HST images of Virgo Cluster galaxies. The
transition objects of Ha\c{s}egan et al. have slightly lower luminosities  ($-12<M_V<-11$) than
the Fornax UCDs, but the brighter ones were found to be significantly
different to globular clusters, both in size and in following galaxy
scaling relations. Ha\c{s}egan et al.\ suggest that ``bona fide'' UCDs
are also distinguished by the presence of dark matter as some of their 
objects have $6<M/L_{V}<9$.

A detailed analysis of the stellar populations of compact objects in
the Fornax Cluster (Mieske et al.\ 2006) has reached quite different
conclusions. They measured spectroscopic metallicities for 26 compact
objects with luminosities spanning both UCDs and globular clusters
and found a break in the distribution at about $M_{V}=-11$. The
more luminous objects have a narrow metallicity distribution with
mean [Fe/H]$= -0.62 \pm 0.05$, whereas the less luminous objects show
a much broader range of metallicity and a significantly lower mean
(0.56 dex lower). There is a break in the size-luminosity relation
for these objects at the same luminosity. Mieske et al.\ note that the
metallicity of the dwarf galaxy nuclei in their sample is
significantly lower than that of the UCDs which, in turn, are better
matched by models of massive young star clusters. They therefore
suggest that the UCDs in the Fornax Cluster are formed as a result of
galaxy mergers, but note that the properties of the Virgo Cluster
UCDs are more consistent with the stripping model.

In this paper we present new imaging and spectroscopic observations
of the Virgo Cluster UCDs listed by Jones et al.\ (2006). These are
analysed along with existing data for the luminous Fornax Cluster UCDs
(Drinkwater et al.\ 2003). In Section~2 we describe the
high-resolution spectroscopic observations, and in Section~3 we
describe the corresponding HST imaging. We
present an overview of our results in Section~4 by comparing the UCDs
with other objects in various projections of the fundamental
plane. In Section~5 we investigate the UCD dynamics in more detail,
calculating their mass-to-light ratios, and in Section~6 we examine
the ages and chemical composition of their stellar populations. Our
main results and conclusions are given in Section~7. We use distance
moduli of 30.92 mag and 31.39 mag for the Virgo and Fornax Clusters
respectively (Freedman et al.\ 2001).

\section{Spectroscopy}

Observations of the six Virgo UCDs, a comparison M87 globular cluster
(Strom417), a comparison dwarf galaxy nucleus (VCC417) and the NGC4486B
galaxy (its central part) were carried out on 2003 April 6--7 with the
Echelle Spectrograph and Imager (ESI) on the Keck II telescope in the
echellette mode (Table~1). A slit width of 0\farcs75 was used,
providing an instrumental resolution of $\approx$ 50 \,km~s$^{-1}$
(FWHM) or $\lambda/\Delta\lambda \approx $ 6,000.  The wavelength
range is $3900-11000\mbox{\AA}$, distributed over 10 echelle orders,
with a dispersion of 11.4\,km~s$^{-1}$~pixel$^{-1}$. The exposure times 
are given in Table~1.  In addition, standard
stars of spectral types in the range between G8III and M0III (Table~2)
were observed for use as templates for the radial velocity and
velocity dispersion measurements and line index calibrators.  The
standard stars were observed in two ways: held centered in the slit,
and also moving perpendicularly across the slit (``smeared'') to
simulate the appearance of extended sources which overfill the slit
and consequently have slightly lower spectral resolution.  The seeing
was a stable 0\farcs8 the first night, and ranged from 0\farcs8 to
1\farcs1 the second.  The first night was photometric; the second had
occasional light cirrus.

The data were reduced using scripts in {\sc IRAF} and IDL specifically
written to handle the ESI data format and idiosyncrasies, but
otherwise are standard procedures for CCD data.  Spectra were
extracted over a 1\farcs5 aperture centered on the peak, taking in
nearly all the light from the UCDs and globulars.  The relative flux
scale was determined using nightly observations of Feige~34.
The S/N for each integration ranges from $\sim 15$ to 25 per pixel,
yielding final S/N of 30 to 50 after coadding the multiple observations of each target.  

The radial velocity and velocity dispersion of our objects were determined using two different techniques: 
the direct-fitting method (as described and implemented by van der Marel 1994) 
and the cross-correlation method (Tonry \& Davis 1979, as implemented in RVSAO/IRAF).
In the direct-fitting method the template star spectrum (Table~2) is broadened with Gaussian functions 
of variable $\sigma$ in velocity space. The resulting set of spectra are then compared 
with the object spectrum. The best-fitting Gaussian function is determined by $\chi^2$ 
minimization in pixel space.
The second method is to cross-correlate the object spectrum with the stellar template spectrum 
to determine the width of the cross-correlation peak and the redshift. The correlation width is used to 
estimate the velocity dispersion by comparison with results from artificially broadening 
the template stars by convolution with Gaussian functions of known width.
This method is less sensitive to the exact match between template and object spectra, than the 
direct-fitting method.

Velocities and velocity dispersions were obtained from the CaT ($8400-8750\mbox{\AA}$) and 
Mgb ($5100-5250\mbox{\AA}$) regions. 
The measured values are given in Tables 3 \& 4 and are consistent, 
within the measurement errors, for the two techniques and the two wavelength regions.
The exception is VUCD3. 
In the following discussion, we use values obtained as the mean of the two wavelength regions and 
the direct-fitting method as it gives smaller measurement errors. 
The adopted velocities and velocity dispersions for the observed objects are shown in Table~5.
Our measurements for NGC4486B are in good agreement with Bender et al. (1992). 

There were almost no differences in velocity and velocity dispersion measurements obtained with  
smeared and unsmeared stellar templates.

The two wavelength regions give significantly different velocity dispersions for VUCD3:  
$\approx$ 49 \,km~s$^{-1}$ for the Mgb region, but only $\approx$ 38 \,km~s$^{-1}$ for the CaT region.  
This discrepancy appears to be real.  
We suspect this may be because the ratio of giant to dwarf population contributions to the continuum 
is varying rapidly in this object compared to the other UCDs. This might be expected because this
object also has a much higher metallicity than the other UCDs (see Section 6 below), so that the 
relative contribution of the higher gravity dwarf stars to the
blue-green (Mgb) region of the spectrum is greater, leading to broader lines in
this region compared to the other UCDs.   The CaT region is
dominated by giant stars, no matter what the metallicity of the UCDs.  
Modeling of this effect could lead to a better understanding of the stellar populations, 
but a more extensive library of stars is required.

\section{Imaging}

We obtained images of the Virgo UCDs in the course of HST snapshot
program 10137.  The data were taken with the Advanced Camera for
Surveys (ACS), High Resolution Channel (HRC) through the F606W and
F814W filters. Exposure times were 870 sec in F606W and 1050 sec in
F814W.  The HRC scale is $0''.025$ pixel$^{-1}$. For the image analysis
we used {\sc
  MultiDrizzle}\footnote{http://stsdas.stsci.edu/multidrizzle}
(mdz) files retrieved from the HST archive.

To measure the total magnitudes, we plotted curves of growth
(integrated magnitude versus circular aperture radius) to find an
aperture radius large enough to enclose all the light from an object.
The instrumental F606W and F814W magnitudes were transformed into
Landolt $V$ and $I$ band following Sirianni et al. (2005).  The
resulting $V$ magnitudes and $V - I$ colors are listed in Table~6.

We have also used HST imaging data (program 8685) for Fornax UCDs and
a dE,N (FCC303), initially presented in Drinkwater et al. (2003), as
one of the aims of this work is to compare Virgo UCDs with Fornax UCDs
and dwarf nuclei. The data consist of 1960 sec exposures taken with
the Space Telescope Imaging Spectrograph (STIS) in unfiltered mode
(50CCD). The STIS has a scale of $0''.0507$ pixel$^{-1}$.  We
re-processed the STIS images with {\sc MultiDrizzle} to ensure that
the data reduction method for the Fornax UCD images was consistent
with that of the Virgo UCD images.  The instrumental AB magnitudes were
transformed into $V$ band using the relation
$50CCD = V + 0.2165 + 0.5831(B-V) - \\ 2.267(B-V)^2 + 2.6626(B-V)^3
-1.128(B-V)^4$ 
(H.Ferguson, private communication; see also Gregg \& Minniti
1997). We
used $B - V$ colors for UCDs and FCC303 from Karick et al. (2003).
The total $V$ magnitudes for Fornax UCDs and FCC303 are presented in
Table~7.

The images of Virgo and Fornax UCDs were modeled using two-dimensional
fitting algorithm GALFIT (Peng et al. 2002) and assuming empirical
King, Sersic and Nuker models for the luminosity profile.

The King profile is characterized by the core radius, $R_c$, and the
tidal radius, $R_t$, and has the following form:
\begin{equation}
I(R) = I_0 \left[\frac{1}{(1+(R/R_c)^2)^{\frac{1}{\alpha}}} - \frac{1}{(1+(R_t/R_c)^2)^{\frac{1}{\alpha}}} \right]^{\alpha},
\end{equation}
where I$_0$ is the central surface brightness.  We tried both the
standard model with $\alpha = 2$ and generalized model with variable
$\alpha$.  The King model provides a good fit to globular cluster
luminosity profiles when the concentration index $c = {\rm
  log}(R_t/R_c) = 0.75 - 1.75$ and to elliptical galaxy luminosity
profiles if $c \ge 2.2$ (Mihalas \& Binney 1981).

The Sersic power law is often used to fit luminosity profiles of
galaxies and has the following form:
\begin{equation}
I(R) = I_{\rm eff} \,\, exp \left[-k \left(\left(\frac{R}{R_{\rm eff}}\right)^{\frac{1}{n}} - 1 \right) \right],
\end{equation}
where $R_{\rm eff}$ is the half-light (effective) radius, $I_{\rm
  eff}$ is surface brightness at the effective radius, n is the
concentration parameter (n=4 for de Vaucouleurs profile and n=1 for
exponential profile) and k is a constant which depends on n.

The Nuker law is as follows:
\begin{equation}
I(R) = I_b \,\, 2^{\frac{\beta - \gamma}{\alpha}} \, \left(\frac{R}{R_b} \right)^{- \gamma} \, 
\left[1 + \left( \frac{R}{R_b} \right)^{\alpha} \right]^{\frac{\gamma - \beta}{\alpha}}. 
\end{equation}
It is a double power law, where $\beta$ is the outer power law slope,
$\gamma$ is the inner slope, and $\alpha$ controls the sharpness of
the transition (at the ``break'' radius $R_b$) from the inner to the
outer region. $I_b = I(R_b)$.  The Nuker law was introduced by Lauer
et al. (1995) to fit galaxy centers.

GALFIT convolves the analytic profile with the PSF and determines the
best-fitting model parameters by minimizing residuals between the
model and original {\it two-dimensional} image.

We derived artificial PSFs for ACS/HRC images in the F606W and F814W
filters using the {\sc TinyTim}
software\footnote{http://www.stsci.edu/software/tinytim} and {\sc
  MultiDrizzle} as follows. First we generated ACS/HRC PSFs with {\sc
  TinyTim}; these include all the distortions, so they represent the
PSF in raw images.  We then implanted these PSFs in empty distorted
images (flt files), {\em at the location of each target observed}, and
passed them through {\sc MultiDrizzle} using the same parameters as
were used for the data.  This produces model PSFs that are processed
the same way as the real data.  For the STIS imaging, the generation
of the PSF is more straight-forward and is achieved through a single
pass of {\sc TinyTim}.

The quality of the GALFIT model fits in the inner regions of each
object is shown in Figures 1 \& 2. These Figures present residual maps
after subtracting the (PSF-convolved) model from each object. The
quality of the models in the outer regions is better shown in Figures
3 \& 4. For these figures we used the ELLIPSE task in IRAF to produce
{\it one-dimensional} surface brightness profiles for the objects and
(PSF-convolved) GALFIT models.

The $\chi_{\nu}^2$ values of the fits (see Peng et al. 2002) are shown
in Tables 8 \& 9. We use $\chi_{\nu}^2$ values to choose the {\it best
  model} for each object (see the last row of Tables 8 \& 9).

From Figures 1-4 and the $\chi_{\nu}^2$ values {\it in both filters} we can
see that the {\it Virgo} UCDs are poorly fitted with the {\it standard
  King model} ($\alpha = 2$), but are very well fitted with the {\it
  Nuker law} -- a double power law (except VUCD7 which requires a
two-component model).  King models predict a truncation radius, beyond
which stars are stripped from the cluster by the galactic tidal
field. {\it None of the UCDs show this tidal cutoff down to our
  limiting surface brightness.}  Also, a main feature of King models
is their central cores---the regions of constant surface
brightness. We have found from Nuker models that only one UCD has a
flat core (VUCD5), all other UCDs have shallow or steep cusps in the
center (Table~8).

{\it Generalized King models} (with a variable $\alpha$ parameter)
provide better fits to the data than standard King models.  The
parameter $\alpha$ controls both the slope of the profile, and the
transition from the main body into $R_t$.  When we relax the $\alpha$
parameter, we can better fit the extended outer parts of the UCDs.
However, King models have a tidal cutoff and do not fit the data as well
as untruncated Nuker models do.

To fit a standard King model to the F606W data for VUCD3, we had to
fix $R_c$ at 1 pixel (1.9 pc).  There is no convergency if we leave
$R_c$ as a free parameter (perhaps because $R_c$ becomes too small, 
$< < $ 1 pixel).
The lack of a good King fit to the F606W data for VUCD3 is 
seen in Figure~3.  We also failed to fit any of the King models to the
F814W data for this object. There was no convergency with
either $R_c$ fixed or relaxed.

{\it Sersic models} do not provide good fits to the Virgo UCDs
either. In the majority of cases, the model profiles drop faster than the
data. The fit seems good in the case of VUCD3, but this object has a
very steep profile (a very high index n).  Models with a Sersic index
n $>$ 2 become very sensitive to the sky determination, flatfielding,
the accuracy of the PSF being used, and how well the assumed profile
agrees with the data (see on-line GALFIT manual\footnote{http://zwicky.as.arizona.edu/$\sim$cyp/work/galfit/TFAQ.html}).  
The more centrally concentrated a galaxy
profile is (the larger the Sersic index n is), the more extended outer
wing it has.  Because of this behavior, if (for example) a profile
already has a high intrinsic index n, a small underestimation of the
background can make n even higher and can cause
large errors in the magnitude and size.
When fitting VUCD3 data in the two passbands, we obtained quite
different parameters: e.g.\ the difference in $R_e$ was 39\,\%.
This could be due to the above reasons.

We conclude that the Nuker law appears to be the best model for all
Virgo UCDs except VUCD7, in which case the King core + Sersic halo model
is the best one.

The situation is different for the {\it Fornax UCDs}. There is no
universal model for them.  Two objects (UCD3 and UCD5) are best fitted
by two-component models.  The other three have very steep profiles
which are hard to model.
In the case of UCD1 the best model is Sersic. The best model for UCD2
is the generalized King law and in the case of UCD4
it is the Nuker law.

The residual maps of Fornax UCD3 reveal faint structure to the
North-West of the core. This is very likely a background spiral galaxy
along the line of sight. This structure affected the profile fits to
UCD3 to the extent that the two-component models gave very inconsitent
results. For this object we therefore restricted the model fits to the
bottom half of the image. The model parameters for UCD3 in all the
tables are from these fits to half the images, although we show the
images and plots in Figures~2 \& 4 for the whole-image fits to
reveal the background object.

The UCD structural parameters obtained from the fits are summarized in
Tables 8 \& 9.  The Nuker fits gave us the outer power law slope $\beta$,
the inner slope $\gamma$, $\alpha$ parameter, the ``break'' radius
$R_b$, the surface brightness of the profile at the ``break'' radius
$\mu_V(R_b)$ and ellipticity $\epsilon = 1 - b/a$ (b/a is the
minor-to-major axial ratio).  From Sersic models, we obtained the
half-light radius $R_{\rm eff}$, Sersic index n, ellipticity
$\epsilon$ and the integrated magnitude $m_{V,tot}$.  The King fits gave
us the core radius $R_c$, tidal radius $R_t$, concentration parameter $c =
{\rm log}(R_t/R_c)$, central surface brightness $\mu_{0,V}$ and
ellipticity $\epsilon$. All the GALFIT models assume a constant
ellipticity which is fitted at the same time as the other parameters. 
The $R_{\rm eff}$ values and the integrated magnitudes
for the King and Nuker models were obtained via numerical integration of
the luminosity profiles (defined by formulae 1 and 3).  We also
obtained half-light radii from the observational data directly
($R_{e,obs}$).  In this case the data were not PSF-deconvolved and the
half-light radii may be overestimated.  The structural parameters for
Virgo UCDs were averaged between the two filters.

For further analysis we had to decide what $R_{\rm eff}$ to choose for
our objects.  We could not take just the best model $R_{\rm eff}$
because, for example, in the case of Virgo UCDs the best model is
Nuker. For all the Nuker models in Tables 8 \& 9 (except for VUCD1 \&
VUCD5) the total volume under the profile (the integrated luminosity)
converges very slowly. It makes $R_{\rm eff}$ estimations uncertain
and results in unphysically large $R_{\rm eff}$ values. This is why we
chose generalized King $R_{\rm eff}$ values. The generalized King models
fit the Virgo UCD data better than standard King and Sersic models,
and they are finite in extent. In addition to this, generalized King
$R_{\rm eff}$ values are consistent with the observational half-light radii
($R_{e,obs}$).

In the case of the Fornax UCDs we also chose the generalized King $R_{\rm
  eff}$ for one-component UCDs and the King $R_{\rm eff}$ for the cores of
the two-component objects. The Sersic fits seem unreliable as all of
the models have very high index n (see above).  Nuker models are not
good for $R_{\rm eff}$ estimation because of the reason explained
above.

We list the best values of the various parameters of the Virgo and
Fornax objects in Table~10. These are used in the following section
for analysis of the scaling relations. In Table~10 we quote three
values of effective radius $R_{\rm eff}$ in the case of two-component
objects (VUCD7, UCD3, UCD5 \& FCC303): the total, only the core, and
only the halo. These were obtained via numerical integration of the
model luminosity profiles.  Table~10 also contains the total V band
apparent magnitude $m_V$, the mean surface brightness within the
effective radius $\left<\mu_V\right>_{\rm eff}$ and ellipticity
$\epsilon$.  The $m_V$ values are observational values copied from
Tables 6 \& 7 except for the core and halo magnitudes of the
two-component objects, which are model values taken from Tables 8 \& 9
(King+Sersic models). The $\left<\mu_V\right>_{\rm eff}$ values were
derived from $R_{\rm eff}$ and $m_V$ as follows:
\begin{equation}
\left<\mu_V\right>_{\rm eff} \, = m_V + 5\,{\rm log}\,R_{\rm eff} +
1.995 \, ,
\end{equation}
where $R_{\rm eff}$ is measured in arcsec. The $\epsilon$ figures are
the best model values from Tables 8 \& 9.

The ellipticities of the UCDs given in Table~10 show that some of the
objects are significantly non-circular; the maximum (core) ellipticity
is 0.24 for Fornax UCD5. 
We have compared the distribution of UCD ellipticities with those
reported for globular clusters in NGC~5128 (Harris et al.\ 2002) and
the Milky Way (Harris 1996; online catalogue version of 2003 February). 
The two-sample Kolmogorov-Smirnov test indicates that
the UCD ellipticities are consistent with both the MW GC distribution---the 
significance level\footnote{By significance level we mean the
  (percentage) probability that the K-S test statistic is as large as
  measured for the null hypothesis that the data sets are drawn from
  the same distribution. Small values of the significance indicate
  that the distributions differ significantly.}
is 53\% or 66\% (depending on the ellipticity used for 
two-component UCDs: core or halo)---and the NGC~5128 GC distribution---the significance 
level is 38\% or 63\%. The Wilcoxon test gives similar results: 
the UCD ellipticities are consistent with the MW GC distribution 
at the 42\% or 36\% significance level and with the NGC~5128 GC distribution at the 27\% 
or 46\% significance level.

\section{Fundamental plane relations}

In this section we compare UCDs with globular clusters and galaxies by
their position in both the luminosity--velocity dispersion plane and
$\kappa$-space (the fundamental plane as defined by Bender et
al. 1992).

First we revise the luminosity-velocity dispersion correlation for
UCDs and other types of stellar systems proposed in Drinkwater et
al. (2003). Figure~5 represents the absolute V magnitude vs. central
velocity dispersion relation for Fornax and Virgo UCDs, globular
clusters (including the most massive and luminous ones: G1 in M31, the
Galactic globular cluster $\omega$ Cen, and NGC5128 massive GCs) and
galaxies.  The UCD data were obtained in the present work (Tables 10 \& 12), 
except central velocity dispersions for Fornax UCDs, which were taken from 
Hilker et al. (2006). The velocity dispersion for Fornax
UCD1 was derived from CaT region using the cross-correlation method
and only one stellar template (G6/G8IIIw type)\footnote{The UCD1
  velocity dispersion was measured and used by Drinkwater et
  al. (2003).}.
The data for globular clusters are from: M31 GCs -- Djorgovski
et al. (1997) and references therein; G1: Djorgovski et al. (1997) and
references therein, except half-light radius (required for Figure~6),
which was taken from Barmby et al. (2002); Milky Way GCs, $\omega$
Cen, LMC and SMC GCs (most of which have old ages) -- McLaughlin \&
van der Marel 2005 (photometry is based on Wilson models); NGC5128 GCs
-- Martini \& Ho (2004); Strom417: spectroscopy -- this work (Table~12),
photometry -- Ha\c{s}egan et al. (2005). We also plot ``dwarf-globular
transition objects'' from Ha\c{s}egan et al. (2005).  Data for
galaxies were obtained from: giant ellipticals -- Faber et al. (1989);
NGC4486B: spectroscopy -- our data, Bender et al. (1992), photometry
-- Faber et al. (1989); the compact elliptical galaxy M32 -- Faber et
al. (1989), Bender et al. (1992); dE,Ns and dwarf nuclei -- Geha et
al. (2002, 2003); FCC303: photometry -- this work, spectroscopy -- Hilker et al. (2006); 
VCC1407: velocity dispersion -- this work,
magnitude -- NED. All the $M_V$ magnitudes were dereddened and all the
velocity dispersions for GCs and UCDs were aperture-corrected to give
standard values (the central velocity dispersions) used in the
literature for the comparison with Galactic GCs. 

From Figure~5 we can see that there is no gap between bright GCs and UCDs
and that the Virgo UCDs have velocity dispersions and luminosities similar
to the Fornax UCDs. 
VUCD3/Strom547 and Strom417 were previously considered to be M87 GCs (Strom et al. 1981). 
These are the two brightest M87 GCs according to Hanes et al. (2001) list.  
Now we see that these GCs lie in the same part of
$M_V - \sigma_0$ plane as UCDs and can be equally considered as UCDs.
The UCDs, along with transition objects of Ha\c{s}egan et al., appear to follow approximately the same
relation between luminosity and velocity dispersion as old globular
clusters.
To make a firm conclusion if the UCDs lie on the extrapolation of the GC relation or not, 
more data on the velocity dispersions for bright GCs (such as NGC5128
and M31 globulars) or fainter UCDs (e.g. recently discovered in the
Fornax Cluster, Drinkwater et al. 2004) are required. 
There is an overlap in luminosities and velocity
dispersions of the dE,N nuclei and the properties of bright GCs, transition objects of Ha\c{s}egan et al.
and UCDs, which is consistent with the stripping hypothesis for GC, 
transition object and UCD formation.

Next we consider the location of UCDs relative to other stellar
systems in the $\kappa$-space (Figure~6).
The $\kappa$-space is a space in which axes are combinations of three
observable parameters (central velocity dispersion, effective radius
and mean intensity inside effective radius) into physically meaningful
parameters.  The $\kappa$-space parameters as defined by Bender et
al. (1992) are as follows:
\begin{equation}
\kappa_1 \equiv (\log \sigma_0^{2} + \log R_{\rm eff}) / \sqrt2 \, ,
\end{equation}
\begin{equation}
\kappa_2 \equiv (\log \sigma_0^{2} + 2 \log I_{\rm eff} - \log R_{\rm
eff})/\sqrt6 \, ,
\end{equation}
\begin{equation}
\kappa_3 \equiv (\log \sigma_0^{2} - \log I_{\rm eff} - \log R_{\rm
eff})/\sqrt3 \, ,
\end{equation}
\noindent
where $\sigma_0$ is the central velocity dispersion in ${\rm
  km~s}^{-1}$, $R_{\rm eff}$ is the half-light radius in kpc, $I_{\rm
  eff}$ is the mean intensity inside $R_{\rm eff}$, defined as
$10^{-0.4(<\mu_V>_{\rm eff}-26.42)}$ and measured in V-band solar
luminosities pc$^{-2}$.
The $\kappa$ variables have the following physical meanings:
$\kappa_1$ is proportional to the logarithm of the mass, $\kappa_2$ is
proportional to the logarithm of the surface brightness cubed times
the mass-to-light ratio, and $\kappa_3$ is proportional to the logarithm
of the mass-to-light ratio.  

As expected, objects are distributed more widely in the
$\kappa_1$-$\kappa_2$ plane, the face-on projection of the fundamental
plane, than in the $\kappa_1$ - $\kappa_3$, the edge-on projection.
In the $\kappa_1$ - $\kappa_3$ plane we show that the UCDs lie on the same
tight correlation between mass and mass-to-light ratio as the bright
GCs and transition objects of Ha\c{s}egan et al., but the fainter GCs ($\kappa_1<0$) show little if any
correlation in this plane. This corresponds approximately to a mass of
$10^6$ M$_\odot$ at which Ha\c{s}egan et al. (2005) also find a turn-over
in scaling relations for GCs and other low-mass systems in other projections of the
fundamental plane. Consistent with previous discussions (e.g.
Burstein et al. 1997) we find that this relation does not intersect
that of giant elliptical galaxies in the same $\kappa_1$ - $\kappa_3$
plane.

By contrast, in the $\kappa_1$ - $\kappa_2$ plane we find that the UCDs
are clearly not on the main GC relation (as defined by the Milky Way
and M31 GCs). The UCDs lie in a region away from this sequence in the
direction of increasing mass ($\kappa_1$). The NGC5128 GCs also
lie off the main GC relation between it and the UCDs. 
We must note here that
the UCDs are from magnitude-limited samples (equivalent to
$0.6<\kappa_1$), so these data do not actually provide evidence for a
gap between UCDs and GCs. There is a relatively empty region in the plane at 
($\kappa_1,\kappa_2$) values of around (0,3.5--4.5) at masses intermediate
between GCs and UCDs but low $\kappa_2$ values. According to Bastian et al. (2006), young
massive star clusters, if evolved to ages of 10 Gyr, would occupy this
region of the fundamental plane. Bastian et al.\ note that not many of
these are expected to survive to such ages, but even those not
disrupted would be unlikely to appear in the existing observational
data sets due to the following selection effects. First, there are no
UCDs in this region as our UCD samples are limited to higher masses
(luminosities strictly speaking). Secondly, GCs in this region would
have high mass and presumably would only be found in external galaxies
like the NGC5128 objects, but they would have much lower surface
brightness values, so it would not be possible to measure their
velocity dispersions even if they could be detected.

It is interesting to note that the more massive objects
($0<\kappa_1<1.5$) show the opposite correlation between $\kappa_1$
and $\kappa_2$ to the lower-mass systems in Figure~6. Since $\kappa_3$
($M/L$) is increasing for the high-mass objects, it must be $I_e^3$
that is decreasing for these objects. This agrees with the mass-size
relation observed for objects in this mass range (e.g.\ Kissler-Patig
et al.\ 2006 and Ha\c{s}egan et al.\ 2005).

In both the fundamental plane projections we also show the parameters
of the nuclei of a sample of nucleated dwarf elliptical galaxies in
the Virgo Cluster (Geha et al. 2002). The nuclei with similar masses
to the UCDs lie in the same region of both plots as the UCDs. This is
consistent with the threshing hypothesis for UCD formation from
disrupted dwarf elliptical galaxies. The UCDs are mostly well-separated from
complete dwarf elliptical galaxies in the fundamental plane; the
closest galaxy to UCDs is M32, the prototype ``compact elliptical''
galaxy which is also thought to have formed through a disruptive process 
(e.g. Choi et al. 2002).

\section{Masses and Mass-to-Light Ratios}

In this section we estimate the masses of the UCDs using dynamical
models.  The masses and mass-to-light ratios of the UCDs are important
physical parameters for the understanding of their origin. In
particular, the mass-to-light ratio ($M/L$) is an indicator for
possibly existing dark matter and/or violation of dynamical
equilibrium or isotropy of stellar orbits.  If UCDs were the
counterparts of globular clusters, one would expect $M/L$ values as predicted
by simple stellar population models (e.g. Bruzual \& Charlot 2003,
Maraston 2005). If UCDs were of ``galaxian origin''---formed in dark
matter halos---they might still be dominated by dark matter and show
high $M/L$ values. Mass-to-light ratios larger than expected
from simple stellar populations can, however, also be caused by
objects that are out of dynamical equilibrium, e.g.\ tidally disturbed
stellar systems (Fellhauer \& Kroupa 2006).

\subsection{Method}

The UCD masses were estimated from the measured velocity dispersions
and their structural parameters. We showed in Section~3 that the UCD
light profiles can be fitted by various functions. A simple King
profile often is not the best choice to represent UCD surface
brightness profiles. However, most mass estimators available in the
literature are based on the assumption of a King profile.  In order to
be not restricted to King profiles, we used a more general approach,
using software developed by H. Baumgardt (see Hilker et al.\ 2006).

The first stage is to deproject the observed density profile (either
King, generalized King, Sersic, Nuker, or King+Sersic), calculate its
distribution function $f(E)$ under the assumption of spherical
symmetry and an underlying isotropic velocity distribution, and finally
create an $N$-body representation of the UCD. It is assumed that mass
follows light (e.g.\ mass segregation is neglected). Besides the
projected profile parameters, the total mass of the stellar system and
the number of test particles are needed. The resulting model is a list
of $x$, $y$ and $z$ positions and $v_x$, $v_y$ and $v_z$ velocities
for all particles that correspond to the specified structural
parameters and the given mass. From this model, the central as
well as global projected velocity dispersion can be calculated. The
projected half-mass radii for all the models were also derived and
were in very good agreement with the half-light radius values in
Tables 8 \& 9.

In the second stage, the velocity dispersion seen by an observer is
simulated.  In order to do this, all test particles are convolved with
a Gaussian whose full-width half-maximum (FWHM) corresponds to the
observed seeing. The fraction of the light (Gaussian) falling into the
slit at the projected distance of the observed object (the size of the
slit in arcsec and the distance to the object in Mpc are input
parameters) is then calculated.  These fractions are used as weighting
factors for the velocities.  The weighted velocities of all particles
whose ``light'' falls into the slit region are then used to calculate
the mimicked observed velocity dispersion $\sigma_{\rm mod}$.

The total ``true'' mass of the modelled object, $M_{\rm true}$, that
corresponds to the observed velocity dispersion, $\sigma_{\rm obs}$ is
not known a priori. One has to start with a first guess of the total
mass, $M_{\rm guess}$, from which the ``true'' mass can be calculated
as $M_{\rm true} = M_{\rm guess}\cdot (\sigma_{\rm obs}/\sigma_{\rm
  mod})^2$.

In the case of the Nuker and Sersic functions, the models were
truncated at large radii to avoid the unphysical infinite extensions
of UCD light profiles.  The truncation radius of the Nuker model was
fixed to 2 kpc. The truncation radius of the Sersic model was set to
20 times the effective radius, thus ranging between a few hundred
parsecs and a few kiloparsecs for the UCDs in our sample. The true
tidal radii of the UCDs depend on their distances to the cluster
center $R_G$ and the enclosed mass $m_c$ of the potential they are
living in. They can be estimated by the formula: $r_t =
(G\,m_c\,/\,2\,v_{\rm circ}^2)^{1/3}\,R_G^{2/3}$, where $v_{\rm circ}$
is the circular velocity of the cluster potential and $G$ is the
gravitational constant. The estimated tidal radii of the UCDs range
between 1 and 4 kpc, thus justifying the chosen truncation radii.

A more detailed description of the mass determination process and mass values for 
Fornax UCDs are presented in Hilker et al. (2006). 
In this paper the discussion will be focused on the Virgo UCDs.

\subsection{Uncertainty analysis}

The uncertainties of the modelled masses were estimated from the
uncertainty in the observed velocity dispersion, $\Delta\sigma$ and
the differences of the model parameters for surface brightness
profiles in the $V$ and $I$ band.  On one hand,
maximum and minimum model masses were simulated that correspond to the
observed velocity dispersion of $(\sigma_{\rm obs}+\Delta\sigma)$ and
$(\sigma_{\rm obs}-\Delta\sigma)$, respectively. The average of the
differences $(M_{\rm max}-M_{\rm true})$ and $(M_{\rm true}-M_{\rm
  min})$ defines the first mass uncertainty. On the other hand, models
were created that simulate $\sigma_{\rm obs}$ from the profile
parameters in $V$ and $I$ separately as well as from combinations of
their parameters (i.e. $R_c$ of $V$ and $R_t$ of $I$) to mimic the
uncertainties in the profile parameters. The maximum and minimum mass
deviations from $M_{\rm true}$ define the second uncertainty. Both
uncertainties then were summed to derive the total mass uncertainty.

The uncertainties for the mass-to-light ratios were propagated from
the mass uncertainties and the uncertainty in the luminosity (assumed
to be 0.05 mag in the absolute magnitude). The luminosities were
derived from the apparent $V$ magnitudes given in Table~10, the
distance to the Virgo Cluster as mentioned in the introduction
and a solar absolute $V$ magnitude of $M_{V,\odot} = 4.85$ mag.

The uncertainty of the central velocity dispersion was estimated from
the observational uncertainty plus the scatter of modelled velocity
dispersions in annuli of 0.5 parsec within the central 5 pc for each
object. The uncertainty of the global velocity dispersion is the sum
of the observational uncertainty and the uncertainties as propagated
from the mass uncertainty of the profile fitting parameters.

\subsection{Results}

The results of the modelled object masses and velocity dispersions for
the King, generalized King, Sersic, Nuker and King+Sersic functions
are presented in Table~11. These results are based on the surface
brightness profile parameters in Table~8 and the observed
velocity dispersions in Table~5. For
Strom417 we used surface brightness profile parameters obtained by
Ha\c{s}egan et al. (2005) from King model fits. 

The masses and $M/L$ values of the different models in general agree
with each other within the uncertainties. On average, the masses of
the Nuker profile models are slightly higher than those derived from the
other profiles, whereas the masses of the Sersic profile models are on the
low side.  As discussed in Section 3, the generalized King models give
the most stable estimates in the case of one-component profile fits.
Therefore, the masses from the generalized King models and from the
King core plus Sersic halo models in the case of VUCD7 were adopted for 
further analyses (see Table~12).

The Virgo UCDs have masses and mass-to-light ratios in the range
$M \approx 2-9 \times 10^7M_\odot$ and $M/L \approx 3-5 \,
M_\odot/L_{\odot,V}$. The Fornax UCD masses and mass-to-light ratios
are in the same range (Hilker et al. 2006).

In order to compare our results with other dynamical mass estimators
we calculated masses using the King mass estimator (e.g. Queloz et
al. 1995) and the virial mass estimator (Spitzer 1987). Both methods
assume a constant $M/L$ ratio as function of radius, an isotropic
velocity distribution and that the object is in virial equilibrium.  
We consider the UCDs are in virial equilibrium as their ages 
(estimated in the following section) are much greater than their crossing times 
($T_{cr} \sim R_{\rm eff}/\sigma \approx 0.4-4$ Myr).

The King mass estimator takes the form:
\begin{equation}
M_K = \frac{9}{2\pi G}\frac{\mu \, R_c \, \sigma_0^2}{\alpha \, p} \,\, ,
\end{equation}
where $\sigma_0$ is the central projected velocity dispersion (from
Table~11, standard King model), $R_c$ is the core radius (from Tables
8 \& 9, standard King model), $G$ is the gravitational constant and
$\mu$, $\alpha$ and $p$ are constants which depend on the
concentration, $c$, and are tabulated in King (1966) and Peterson \&
King (1975). $\mu$, $\alpha$ and $p$ are the mean of the two passbands
for Virgo UCDs, except for VUCD6. In the case of VUCD6 we used $V$
passband only, $I$ passband gives unreasonably high mass estimate.
$R_c$, $\mu$, $\alpha$ and $p$ for Strom417 were taken from Ha\c{s}egan et
al. (2005).

The virial mass estimator is as follows:
\begin{equation}
M_{\rm vir} \approx 9.75 \frac{R_{\rm eff} \, \sigma^2}{G} \,\, ,
\end{equation}
where $\sigma$ is the global projected velocity dispersion (from Table
12) and $R_{\rm eff}$ is the half-light radius (from Table
10). $R_{\rm eff}$ for Strom417 was taken from Ha\c{s}egan et al. (2005).

The results of these two mass estimators are given in Table~13.  
The uncertainties of the calculated masses were propagated from the
uncertainties in $R_c$ and $\sigma_0$ for the King mass estimator and
the uncertainties in $R_{\rm eff}$ and $\sigma$ for the virial mass
estimator where the uncertainty in $R_c$ was estimated from the
difference of $R_c$ in the $V$ and $I$ profile.

Both the King and virial masses are consistent 
with the masses and mass-to-light ratios indicated in Figure~6 
and with the masses derived from the dynamical models (Table~12). 
The exception is VUCD7 for
which the virial mass estimate is 1.8 times larger than the dynamical
model mass. This may be due to the inability of the simple virial
estimator to correctly model the prominent core structure of this
object, but the difference is not significant given the large
uncertainty in the virial mass estimate for this object.

In Table~14 we compare the dynamical $M/L$ estimates for Virgo UCDs
with the predicted stellar mass-to-light ratios of simple stellar
population (SSP) models by Maraston (2005).  To obtain the SSP model
values, we used UCD ages and metallicities derived in the following
section (Figure~9). The uncertainties of the $M/L$ values are based
on the age and metallicity ranges.  The dynamical $M/L$ values are
consistent with the SSP model predictions within the uncertainties for
both Salpeter and Kroupa IMF. It implies that Virgo UCDs do not
require dark matter to explain their mass-to-light ratios. 
This conclusion applies to the central region where we have velocity 
dispersion data covered by our spectroscopic observations. 
An increasing dark matter contribution towards larger radii can not be 
ruled out with the present data.

The mass-to-light ratios of Fornax UCDs are discussed in Hilker et
al. (2006).  

The mass-to-light ratio of Strom417 ($6.6\pm1.5$) is larger than the
$M/L$ value predicted by the SSP models with Kroupa IMF, but is in
agreement (within the uncertainties) with the predictions from the
models with Salpeter IMF. This mass-to-light ratio is consistent with
the high value reported for this object by Ha\c{s}egan et al.\ (2005),
but our result is based on their King model fit to this object. For
this reason our measurement is not an independent confirmation of
their result.

\section{Ages and Chemical Compositions}

In this section we estimate the ages, metallicities, and abundances of
our objects using the Lick/IDS index analysis.  Lick/IDS
absorption-line indices were measured as defined by Worthey et
al. (1994) in the wavelength region $4800-6500\mbox{\AA}$. We could not
use the spectral region $3900-4800\mbox{\AA}$ due to low S/N.  The
line indices H$\beta$, Mgb, Fe5270, and Fe5335 together with
$1\,\sigma$ uncertainties are listed in Table~15.  Before measuring
the indices we shifted the spectra to the rest frame wavelengths.  Since our
spectra have much higher resolution ($\sigma_{ESI} \approx 0.3
\mbox{\AA}$ in the blue wavelength range) than the Lick/IDS system
($\sigma_{Lick} \approx 3.6 \mbox{\AA}$ at the H$\beta$ - Fe5335
wavelength range), we smoothed our spectra to the resolution of the
Lick data with a Gaussian of dispersion $\sqrt{\sigma_{Lick}^2 -
  \sigma_{ESI}^2} \approx \sigma_{Lick}$.  The estimated size of the
smoothing kernel is $\sim$ 17 ESI pixels (Gaussian sigma).  After
broadening, the Lick indices for our objects and for the Lick/IDS
standards (9 stars) were measured as described in Worthey et
al. (1994).  The uncertainties on the indices were calculated
according to Cardiel et al. (1998) based on the noise spectrum of each
galaxy/GC.

To check the agreement between our instrumental system and Lick/IDS system, we calculated the difference between 
measured and published (Worthey et al. 1994) indices for all observations of the 9 calibration stars. 
Figure~7 shows our standard star measurements versus the published values (Worthey et al. 1994).
The mean offsets between our instrumental system and the Lick/IDS system are listed in Table~16.
The index measurements were corrected for the offsets.

The Lick index measurements for the NGC4486B galaxy were also
corrected for the effects of internal velocity dispersion as described
in Davies et al. (1993). UCDs, GCs and dE,Ns have small internal
velocity dispersions compared to the Lick/IDS broadening
function. There is no need to apply velocity dispersion correction for
these objects. 

We have not corrected the dE,N (VCC1407) spectrum for any halo
contribution because the nuclear light dominates in the central 1\farcs5
of our extraction; this object is already on the old envelope and its
indices agree well with the other Virgo dwarf elliptical galaxies
measured by Geha et al. (2003). 

The N4486B galaxy does not have a distinct halo, like the dEs do and 
the problem is the opposite of the dE problem.
When subtracting ``sky'' from along the short slit like ESI ($20''$), one really
subtracts a component of galaxy light from farther out, so then one
ends up subtracting too much stellar light, which then alters the
actual physical extent of what one samples. However, since
ellipticals (especially the likes of N4486B) are very peaked in the
center and have very modest (if any) line strength gradients, the
error in the measured indices is very small.

To translate measured line indices into age and metallicity estimates,
we used the SSP models of Thomas et al. (2003). These
models predict Lick indices for a wide range of ages (1-15 Gyr) and
metallicities ([Z/H] = -2.25,-1.35,-0.33,0.0,+0.35,+0.67 dex), and are
tabulated for several different abundance ratios ([$\alpha$/Fe] =
-0.3,0.0,+0.3,+0.5).

To estimate [$\alpha$/Fe] for the UCDs we plot Mgb (an indicator of $\alpha$--elements) versus $<$Fe$>$ 
(an average of the indices Fe5270 and Fe5335) in Figure~8, overlaid with isochrones and isometallicity lines from 
Thomas et al. (2003). Five UCDs and Strom417 (a GC) have super-solar abundance ratio, 
[$\alpha$/Fe] $\approx$ +0.3 -- +0.5, and one UCD appears to have solar abundances, [$\alpha$/Fe] $\approx$ 0.0.  
The super-solar abundances ([$\alpha$/Fe] $\approx$ +0.3) are typical of old stellar populations 
like globular clusters and elliptical galaxies.
[$\alpha$/Fe] traces the timescale of star formation activity in galaxies.
The majority of $\alpha$-elements is produced rapidly by Type II supernovae, while Fe is produced by Type Ia SNe 
on longer timescales. Supersolar [$\alpha$/Fe] indicates rapid enrichment from Type II supernovae and implies 
that the galaxy/GC has undergone a short burst of star formation activity. The solar and subsolar abundance 
ratios indicate slower chemical enrichment or a more quiescent star formation history (van Zee, Barton \& 
Skillman 2004).

The nuclei of nucleated dwarf ellipticals, taken from Geha et
al. (2003), are also shown in the same plot.  The majority of the dE
nuclei data are consistent with solar [$\alpha$/Fe] abundance ratios, while the
majority of UCDs have super-solar [$\alpha$/Fe] abundances. This provides evidence that the Virgo
UCDs and typical dE,N nuclei are different in that they have different
formation histories.  Our dE,N (VCC1407) lies, however, along the UCD
relation together with two dE,Ns from Geha et al. (2003).

In Figure~9 we show the age-sensitive H$\beta$ index versus the
metallicity sensitive [MgFe]' index ([MgFe]'=$\sqrt{\rm
  Mgb(0.72 \times Fe5270+0.28 \times Fe5335)}$) and compare them with the SSP models
of Thomas et al. (2003).  [MgFe]' is largely independent of
[$\alpha$/Fe] and serves best as a tracer of total metalicity (Thomas
et al. 2003); and H$\beta$ is less [$\alpha$/Fe]-sensitive than other
Lick Balmer line indices (Thomas et al. 2004).  As we can see from the
plot, {\it the Virgo UCDs are old (older than 8 Gyr) and have metallicities between
[Z/H] = -1.35 and +0.35 dex.}  The SSP models
in Figure~9 are shown for the abundance ratio [$\alpha$/Fe] = +0.3.
The conclusion about UCD ages and metallicities remains the same if we
use [$\alpha$/Fe] = 0.0 and +0.5 models.

As a consistency test, we used the Virgo UCD ages and metallicities
derived from Figure~9 and Maraston (2005) SSP models to predict
photometric colors and to compare them with the observed ones.  The
results are summarized in Table~14. The observed colors of Virgo UCDs
are in very good agreement with the colors predicted from the derived
ages and metallicities.

The ages, metallicities and abundances of Virgo UCDs are similar to
those found for GCs in the galaxies M49 and M87 in the Virgo Cluster by
Cohen et al. (2003, 1998).  According to Cohen et al. (2003), the M49 GCs
have metallicities in the range from [Z/H] = -1.3 to +0.5 dex and in
mean are older than 10 Gyr.  The metallicity and age parameters for
M49 and M87 GCs are basically identical.  The GC systems of both of
these galaxies are $\alpha$-enhanced by a factor of about 2 above the
solar value.

We also find that the Virgo UCDs have older integrated stellar
populations on average than the present-day dE,N nuclei. However we
note that UCDs are not distinct from the oldest dE,N nuclei of
Geha et al. (2003) and our dE,N, VCC1407, lies with the UCDs.

The general trend of UCDs to lower metallicities and older ages than
the dE nuclei in Figure~9 is not consistent with the naive threshing
model in which UCDs are identical to the present-day nuclei of dE
galaxies. These results may, however, be consistent with variations of
the threshing hypothesis in which the parent objects are disrupted at
an early time when star formation is still going on and gas is present
(e.g.\ Miske et al. 2006). In this scenario, the stripping selectively
halts the star formation in the stripped objects (UCDs) giving them
the lower metallicities and older ages compared to the nuclei which
continue to form stars. 
However this may not be consistent with the [$\alpha$/Fe] abundances
found in Virgo UCDs: their super-solar abundances imply rapid
enrichment in a short burst of star formation. This seems to be
inconsistent with gas stripping over an extended period, unless the
stripping process caused a very sudden halt to the star formation.

We also measured the near-infrared Ca II triplet (CaT) index for Virgo UCDs 
(as defined in Cenarro et al. 2001) to compare it  
with the metallicity derived from the Lick indices above. 
SSP models predict a strong dependence of the CaT index 
on metallicity for sub-solar metallicities (Vazdekis et al. 2003, Maraston 2005). 
Globular clusters are found to follow the model predictions very well for 
the metallicities typical of Galactic GCs -- up to about a solar metallicity 
(Saglia et al. 2002, Maraston 2005), 
whereas normal and dwarf elliptical galaxies deviate from SSP model predictions 
(Saglia et al. 2002, Michielsen et al. 2003).  
Michielsen et al. (2003) obtained CaT values for a sample of dEs and found that 
four of five dEs with independent metallicity 
estimates have CaT $\sim 8\mbox{\AA}$, which is much higher than expected from their 
low metallicities (-1.5$<$[Z/H]$<$-0.5). 
Saglia et al. (2002) found that the CaT values for bright ellipticals ($7.3\pm1.0\mbox{\AA}$) 
are lower than predicted by SSP models for their ages and metallicities (0.0$<$[Z/H]$<$+0.7).    

The CaT index values for our Virgo objects are listed in Table~17.
The uncertainties were calculated based on the noise spectrum of each object. 
The resolution of the Cenarro et al. stellar library is very close to our spectral
resolution, so no correction for resolution was needed.
The measured indices for stars in common with the Cenarro et al. library showed good
agreement between our instrumental system and Cenarro et al. system.
The CaT index for the NGC4486B galaxy was corrected for the effects of internal velocity dispersion
using a K-type stellar template (see Cenarro et al. 2001).

Figure~10 presents [Z/H] metallicity vs. CaT index for the Virgo UCDs, Strom417 (a GC), VCC1407 (a dE,N)  
and the NGC4486B galaxy. 
The metallicities were derived from the Lick indices (Figure~9).
In Figure~10 we also plot Maraston (2005) SSP model predictions with a Salpeter IMF 
for ages 4, 9 and 15 Gyr. 
It is hard to make any strong conclusions from this figure due to the large uncertainties in the data, 
but it appears that for sub-solar metallicities the Virgo UCDs and Strom417 follow 
the SSP model predictions (within the uncertainties). 
VUCD3 deviates strongly from the model predictions.
It has super-solar metallicity and lies in the plot with NGC4486B and other Es 
from Saglia et al. (2002).  There are no CaT data for GCs at these metallicities available in 
the literature yet, so we can not conclude if VUCD3 is globular-like or not. 
Our dE,N (VCC1407) lies with the UCDs in the [Z/H] vs. CaT plot.
The CaT value for VCC1407 is consistent with the SSP model predictions (within the uncertainties).
However, as we already mentioned above, one of the five dEs studied by Michielsen et al. (2003) also have CaT
in agreement with the model predictions for its metallicity, but all the rest do not. 
Given that globular clusters are known to follow SSP model predictions (for sub-solar metallicities), 
whereas dE and E galaxies do not (Saglia et al. 2002, Michielsen et al. 2003), then Figure~10 
suggests that the Virgo UCDs have CaT indices more like globular clusters than galaxies.

\section{Summary and Conclusions}

In this paper we have presented new imaging and spectroscopic
observations of six Virgo Cluster UCDs (discovery reported by Jones et
al.\ 2006), along with re-analysed data for five Fornax Cluster UCDs
(initially presented by Drinkwater et al.\ 2003).  These are the most
luminous UCDs: $-14<M_V<-12$. The main results of our analysis of
these data are as follows.

\begin{enumerate}

\item From the HST imaging we find that most of the UCDs have shallow
  or steep cusps in their cores; only one UCD has a flat ``King''
  core. We also find that none of the UCDs show tidal cutoffs down to
  our limiting surface brightness. These properties are not consistent
  with the standard King models with flat cores and tidal cutoffs used
  for most globular clusters. However, recent work has shown that GCs
  can have such parameters. Noyola \& Gebhart (2003) obtained inner
  logarithmic slopes of profiles for 28 Galactic GCs and found that
  the slopes span a continious range from zero to 0.6 featuring
  central cusps as well as flat King cores. It is known that young GCs
  can have extended halos (e.g. Elson et al. 1987), but McLaughlin \&
  van der Marel (2005) have now shown that extended halos are a
  generic characteristic of massive GCs---both young and old---in the
  Magellanic Clouds.

\item Fundamental plane projections reveal a) that Virgo UCDs have
  properties similar to Fornax UCDs and b) that UCDs and transition objects of Ha\c{s}egan et al. 
  appear to follow the same relation between
  luminosity and velocity dispersion as old globular clusters.

  In the $\kappa_1$ - $\kappa_3$ plane the UCDs lie on the same tight
  correlation between mass and mass-to-light ratio as the bright GCs
  and transition objects of Ha\c{s}egan et al., but the fainter GCs ($\kappa_1<0$) show little if any
  correlation in this plane. This corresponds to a mass of $\approx
  10^6 M_\odot$ at which Ha\c{s}egan et al. (2005) find a turn-over in
  scaling relations for low-mass systems in other projections of the
  fundamental plane.

  In the $\kappa_1$ - $\kappa_2$ plane the UCDs are not on the main GC
  relation as defined by the MW and M31 GCs, but the available data do not
  provide any evidence for a gap between UCDs and GCs in this plane.

  The dE,N nuclei in the Virgo Cluster with similar masses/luminosities to the UCDs
  lie in the same region of all fundamental plane projections as the
  UCDs.  This is consistent with the threshing hypothesis for UCD
  formation from early-type dwarf galaxies by the removal of low
  surface brightness envelope.

\item The age and metallicity analysis shows that Virgo UCDs are old (older
  than 8 Gyr) and have metallicities ranging from [Z/H] = -1.35 to +0.35 dex.

  The observed colors of Virgo UCDs are in agreement with the colors
  predicted from the derived ages and metallicities.

  Five UCDs and Strom417, a GC, have super-solar abundance ratio,
  [$\alpha$/Fe] $\approx$ +0.3 -- +0.5, and one UCD has solar
  abundance ratio, [$\alpha$/Fe] $\approx$ 0.0.  The super-solar
  [$\alpha$/Fe] abundances are typical of old stellar populations found in globular
  clusters and elliptical galaxies.

  Virgo UCDs and typical present-day dE,N nuclei are different in that they have
  different [$\alpha$/Fe] abundance ratios and, therefore, have
  different formation histories.

  The ages, metallicities and abundances of Virgo UCDs are similar to
  those found for GCs in the galaxies M49 and M87 in the Virgo
  Cluster.

  UCDs generally have lower metallicities and older ages than dE
  nuclei: this is not consistent with the naive threshing model in
  which UCDs are identical to the present-day nuclei of dE galaxies.

  Measurements of the near-IR CaT index suggest that Virgo UCDs have stellar populations 
  more like those found in globular clusters than in dE and E galaxies. 

\item The Virgo and Fornax UCDs all have masses $\approx 2-9 \times
  10^7M_\odot$ and mass-to-light ratios $\approx 3-5 \,
  M_\odot/L_{\odot,V}$.

  UCDs are more massive than transition objects of Ha\c{s}egan et al. 
  and all known GCs for which
  dynamical mass estimates are available.\footnote{The systems with
    dynamical masses are G1: $7-17 \times 10^6M_\odot$ (Meylan et
    al. 2001), $\omega$ Cen: $5 \times 10^6M_\odot$ (Meylan et
    al. 1995), NGC5128 GCs: $1-9 \times 10^6M_\odot$ (Martini \& Ho
    2004), transition objects of Ha\c{s}egan et al. (2005): $0.5-2.5 \times 10^7M_\odot$.}
  
  Although the UCDs are more massive than known globular clusters,
  they are within the theoretical limits for the most massive globular
  clusters formed in galaxies as large as M87 and NGC~1399 (see
  Formula 8 of Kravtsov \& Gnedin 2005)\footnote{The mass of M87
    within 32 kpc is $2.4 \pm 0.6 \times 10^{12} M_{\odot}$ according
    to Wu \& Tremaine (2006) and the mass of NGC1399 within 50 kpc is
    $\approx 2.0 \times 10^{12} M_{\odot}$ as found by Richtler et
    al.\ (2004).}. Recent simulations (Yahagi \& Bekki 2005, Bekki \& Yahagi 2006) confirm
  that globular clusters can escape the potential of these galaxies.

  The UCD masses are close to the estimated masses of some young
  massive GCs (YMGCs) such as NGC7252:W3, NGC7252:W30 and NGC1316:G114
  (Maraston et al. 2004, Bastian et al. 2006), whose origin is
  suggested to be by early mergers of lower mass stellar clusters
  (Kissler-Patig et al. 2006). As these objects evolve they will lose
  mass and their structural parameters will also change (e.g.\
  Fellhauer \& Kroupa 2005). It is not clear that they will still
  have the same masses (and other parameters) as UCDs after the very
  long evolution times demanded by our old age estimates for the Virgo UCDs.

  The dynamical mass-to-light ratios for Virgo UCDs are consistent
  with the simple stellar population model predictions (by Maraston
  2005) within the uncertainties. It implies that Virgo UCDs do not
  require dark matter to explain their mass-to-light ratios.
  This conclusion applies to the central region where we have velocity 
  dispersion data covered by our spectroscopic observations. 
  An increasing dark matter contribution towards larger radii can not be 
  ruled out with the present data.

\end{enumerate}

Note that whilst the structural properties and internal velocity
dispersions have been measured for all the UCDs, the ages and
metallicities (and the interpretation of the mass-to-light ratios) are
limited to the Virgo UCDs for which we have this data.
The high resolution spectra for Fornax UCDs used by Hilker et al. (2006) 
have too low S/N for the Lick index measurements. Also, no Lick standards 
were observed for the calibration onto the Lick/IDS system.
Mieske et al. (2006) present [Fe/H] metallicities and H$\beta$ indices 
for 26 compact objects in Fornax, including UCD2, UCD3 and UCD4 and 
four more objects with luminosities similar to our five Fornax UCDs.
The metallicities derived for Virgo UCDs are total metallicities [Z/H].
[Z/H] and [Fe/H] are related as follows: [Fe/H]=[Z/H]-0.94$\times$[$\alpha$/Fe] 
(Thomas et al. 2003). Assuming Fornax and Virgo UCDs have similar 
mean $\alpha$-abundances, [$\alpha$/Fe]=+0.3, 
the bright Fornax compact objects in Mieske et al. (2006) have mean 
metallicity [Z/H] $\approx$ -0.34, which is similar to the mean 
metallicity for Virgo UCDs.  
The H$\beta$ indices in Mieske et al. (2006) are not calibrated, 
so no reliable conclusions can be drawn, but 
the appearance at least is for the Fornax objects to have higher
H$\beta$ ($2.02 - 2.86 \mbox{\AA}$), hence younger mean ages compared to the Virgo UCDs.  
This suggests a difference in formation time, if not mechanism, for UCDs in the two
galaxy clusters. However, we prefer to refrain from any firm conclusions on  
Fornax UCD origins until accurate age, metallicity and $\alpha$-abundance estimates 
are obtained for them; these should in turn be compared to the
properties of {\em Fornax Cluster} globular clusters and dwarf galaxy
nuclei. In the following discussion, our conclusions focus mainly 
on the Virgo UCDs.  
 
The common feature in all the above results is that our detailed
measurements of the internal UCD properties give values consistent
with observed properties of globular clusters. In all the parameters
we have investigated there is no evidence for any gap between globular
clusters and UCDs.  
The ages, metallicities and abundances of the
Virgo UCDs are similar to those found for GCs in the two brightest
Virgo galaxies (M49 and M87).  This suggests that UCDs and GCs could
have the same formation epoch and the same star formation
history. Theoretical work shows that such massive objects as UCDs
could form in M87 and NCG~1399 and subsequently escape the
host galaxy potential (Kravtsov \& Gnedin 2005; Yahagi \& Bekki 2005, Bekki \& Yahagi 2006).
The surface brightness structure of the UCDs is not different to that
of globular clusters in the Milky Way and Magellanic Clouds. The
mass-to-light ratios of the Virgo UCDs are consistent
with simple stellar populations as in globular clusters.

{\it We therefore conclude that the internal properties of Virgo UCDs 
  are consistent with them being the high-mass/high-luminosity extreme of known 
  globular cluster populations.}

Some of our results, notably the fundamental plane projections are
consistent with the formation of UCDs by the simple removal of the
halo from the nuclei of nucleated dwarf galaxies. However the ages, 
metallicities and abundances for Virgo UCDs are not consistent with this simple
stripping model. It might be consistent with more sophisticated models
of the stripping process that include the effects of gas removal on
the chemical evolution of the nuclei.

As we have shown that the Virgo UCDs are old, we note that definitive
tests of theories of their formation by stripping processes or the
evolution of merger-formed massive star clusters will need to consider
the effects of gas processes (especially gas removal) over these long
timescales.

\acknowledgments

EAE and MJD acknowledge support from the Australian Research Council.
MDG gratefully acknowledges support by the National Science Foundation
under Grant No.~0407445.  Some of the work reported here was done at
the Institute of Geophysics and Planetary Physics, under the auspices
of the U.S. Department of Energy by Lawrence Livermore National
Laboratory under contract No.~W-7405-Eng-48.  
The authors particularly wish to thank Chien Peng (STScI) for very
valuable assistance and discussions concerning the use of his GALFIT
software. We also wish to thank Harry Ferguson (STScI) and Anton
Koekemoer (STScI) for assistance with the HST data analysis, Holger
Baumgardt (University of Bonn) for help with the dynamical calculations and
Richard White (STScI) for allowing us to use his ESI reduction
scripts, which we then modified for our specific purposes.
We are grateful to the referee for very helpful comments and 
suggestions which have improved this paper. 

This work is based on 
data obtained at the W.~M.~Keck Observatory, which
is operated as a scientific partnership among the California Institute
of Technology, the University of California, and the National
Aeronautics and Space Administration.  The Observatory was made
possible by the generous financial support of the W.~M.~Keck Foundation.



{\it Facilities:} \facility{HST (ACS, STIS)}, \facility{Keck:II (ESI)}




\clearpage



\begin{figure}
\epsscale{1.0}
\caption{{\bf Virgo UCD} images in the F606W filter (first column) and residual maps after subtracting 
GALFIT models (PSF convolved) from the UCD images. All the images have a size of $\approx5.''4\times5.''4$ 
($\approx$ 401 pc $\times$ 401 pc).}
\end{figure}

\begin{figure}
\epsscale{1.0}
\caption{{\bf Fornax UCD} images (first column) and residual maps after subtracting GALFIT models 
(PSF convolved) from the UCD images.
 All the images have a size of $\approx6.''6\times6.''6$ ($\approx$ 607 pc $\times$ 607 pc).}
\end{figure}

\begin{figure}
\epsscale{1.0}
\caption{Surface brightness profiles for {\bf Virgo UCDs}, measured in the F606W images. 
The instrumental magnitudes have been transformed into $V$ band. 
The open circles represent UCD profiles; the dashed line represents the best-fitting model convolved 
with the PSF.}
\end{figure}

\begin{figure}
\epsscale{1.0}
\caption{Surface brightness profiles for {\bf Fornax UCDs} and a Fornax dE,N (FCC303). 
The instrumental magnitudes have been transformed into $V$ band. 
The open circles represent UCD profiles; the dashed line represents the best-fitting model convolved 
with the PSF.}
\end{figure}

\begin{figure}
\epsscale{1.00}
\plotone{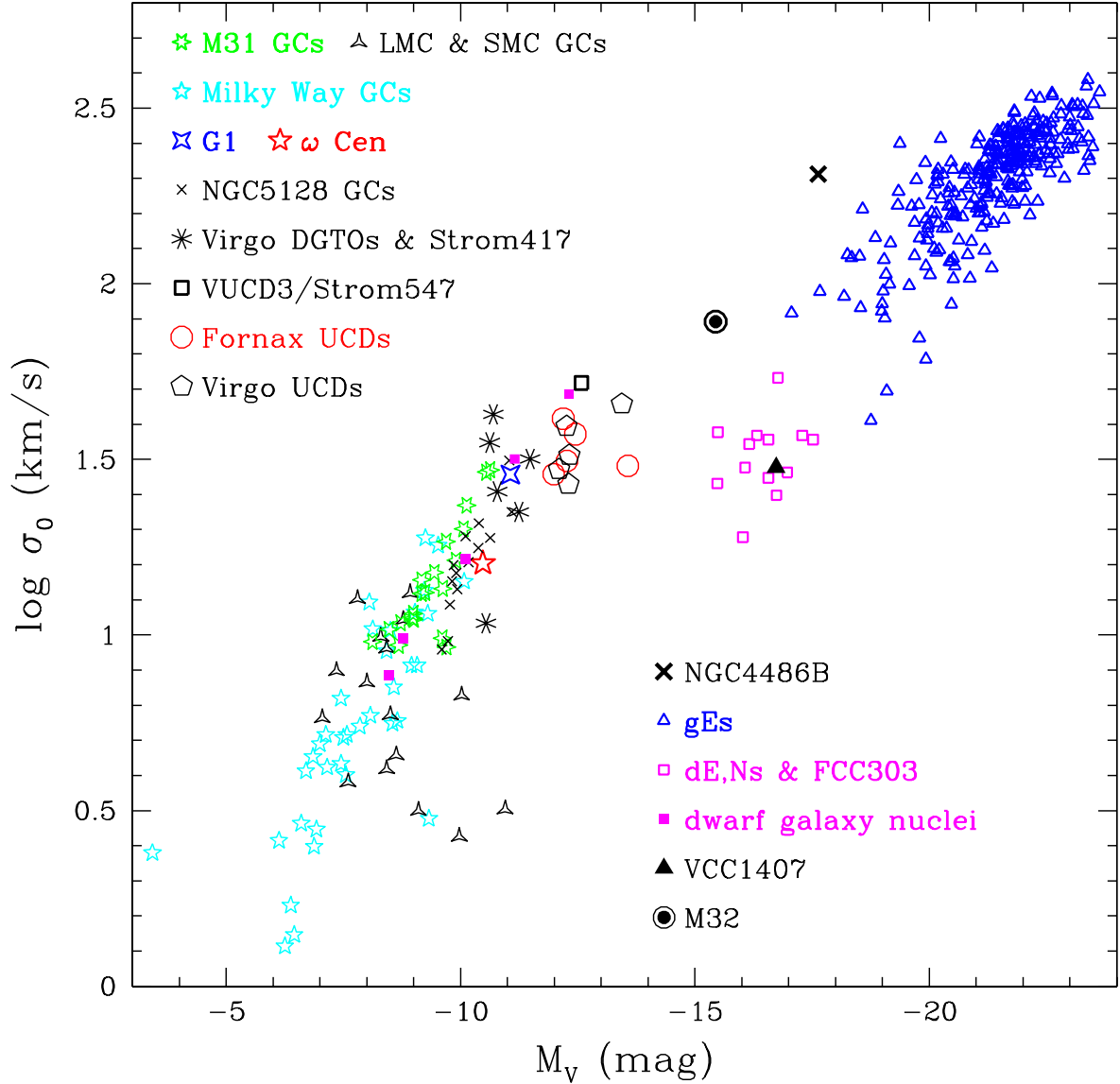}
\caption{Comparison of the internal dynamics of UCDs with globular clusters and galaxies. The sources of 
the data are described in the text.}
\end{figure}

\begin{figure}
\epsscale{1.00}
\plotone{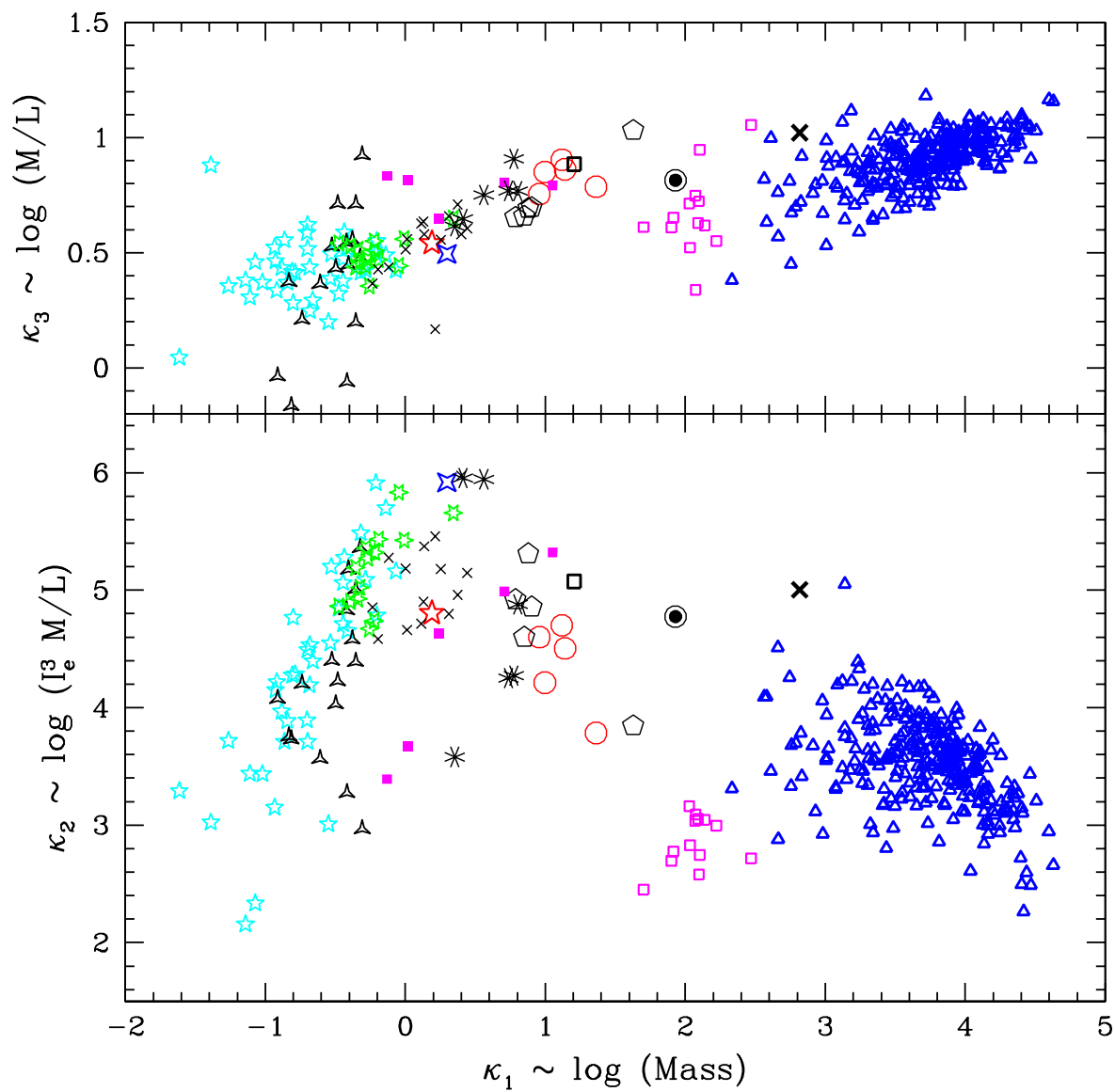}
\caption{The fundamental plane for dynamically hot stellar systems (as defined by Bender et al. 1992): 
top -- edge-on view, bottom -- face-on view. 
Symbols are the same as in Figure~7. The sources of the data are described in the text.}
\end{figure}

\begin{figure}
\epsscale{0.80}
\plotone{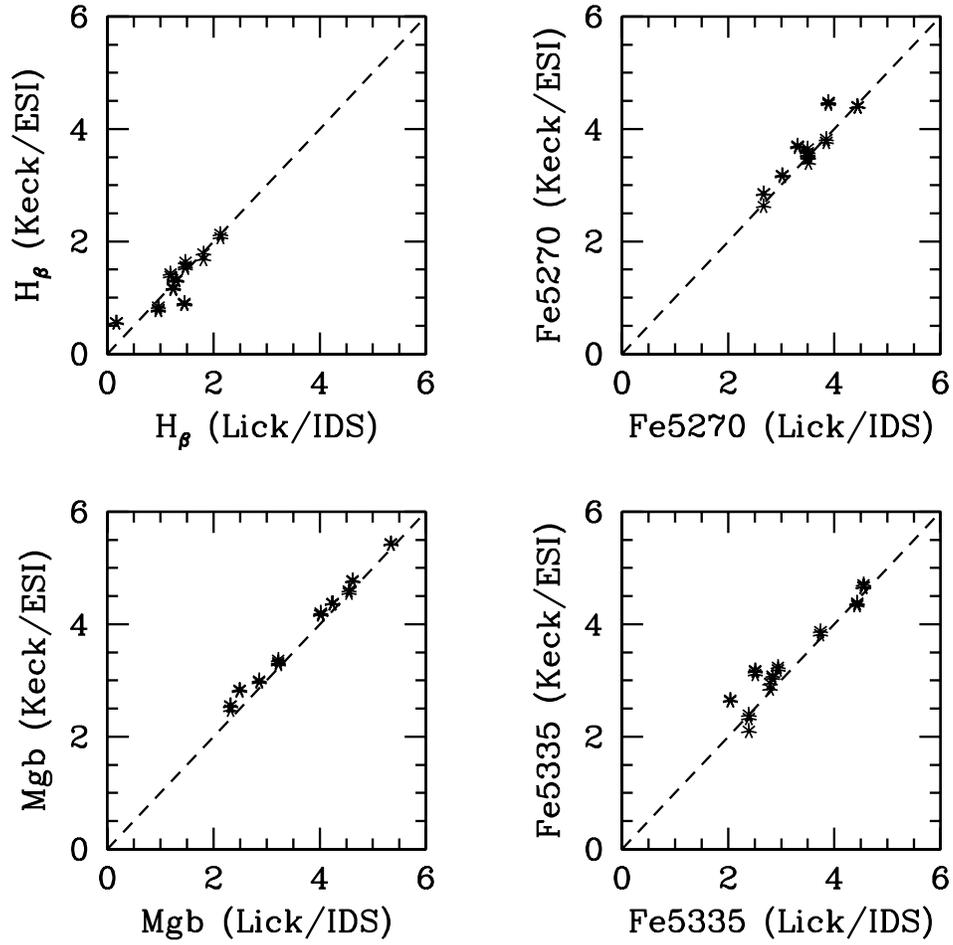}
\caption{Our standard star index measurements versus published values from Worthey et al. (1994).}
\end{figure}

\begin{figure}
\epsscale{0.80}
\plotone{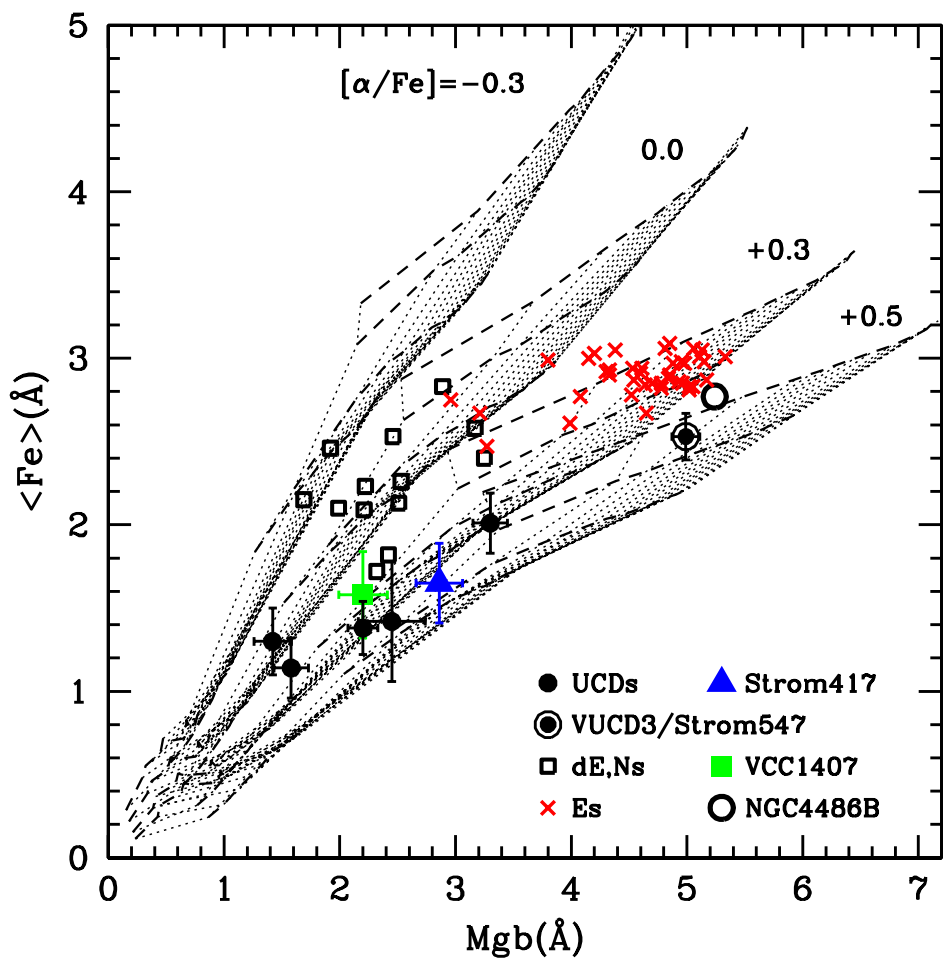}
\caption{Comparison of our data, Mgb versus $<$Fe$>$, with model grids from Thomas et al. (2003).
Open squares are dE,Ns from Geha et al. (2003). Elliptical galaxies from Trager et al. (2000) are shown with 
crosses. Other symboles represent our data.
Thomas et al. (2003) models with variable [$\alpha$/Fe] are plotted for ages of 1-15 Gyr in increments of 1 Gyr 
(dotted lines, from left to right), and metallicities  -2.25,-1.35,-0.33,0.0,+0.35,+0.67 dex (dashed lines, from 
bottom to top).}
\end{figure}

\begin{figure}
\epsscale{0.80}
\plotone{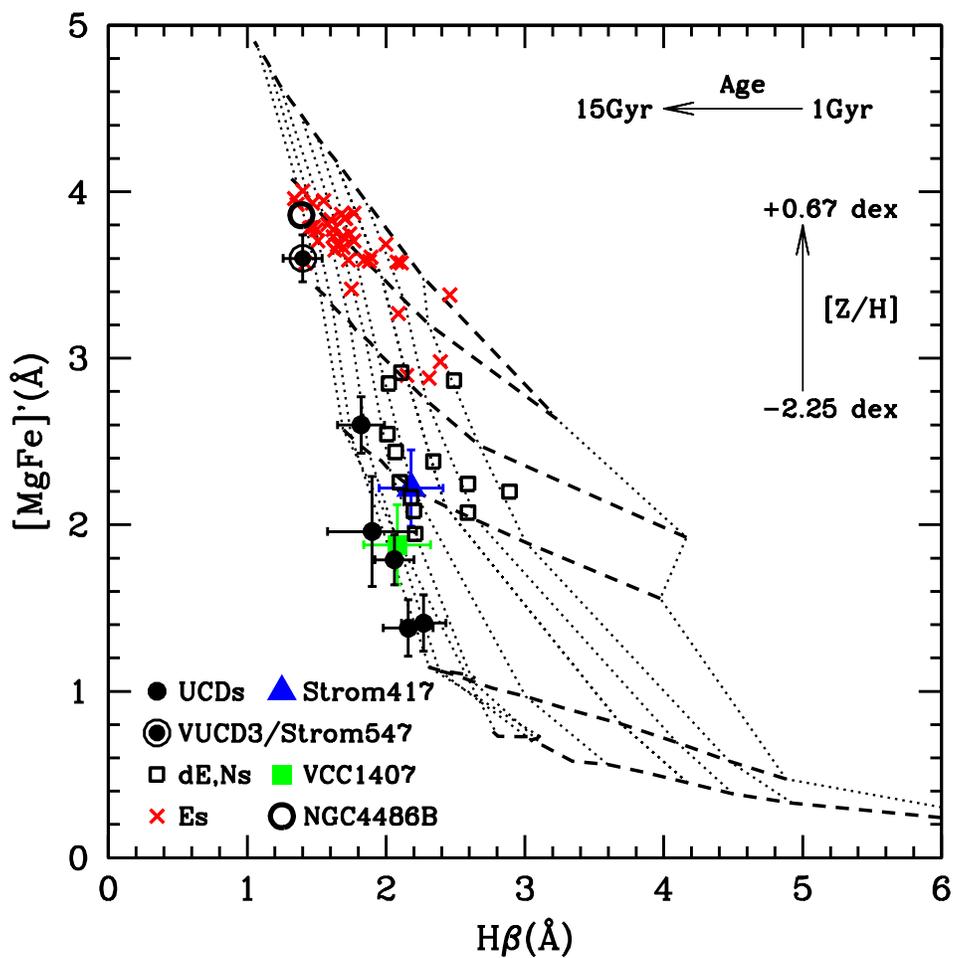}
\caption{Comparison of our data, H$\beta$ versus [MgFe]', with model grids from Thomas et al. (2003).
Open squares are dE,Ns from Geha et al. (2003). Elliptical galaxies from Trager et al. (2000) are shown with 
crosses. Other symboles represent our data.
Thomas et al. (2003) models are plotted for ages 1,2,3,4,6,8,10,12,14 Gyr 
(dotted lines, from right to left), metallicities  -2.25,-1.35,-0.33,0.0,+0.35,+0.67 dex (dashed lines, 
from bottom to top), and [$\alpha$/Fe]=+0.3.}
\end{figure}

\begin{figure}
\epsscale{0.80}
\plotone{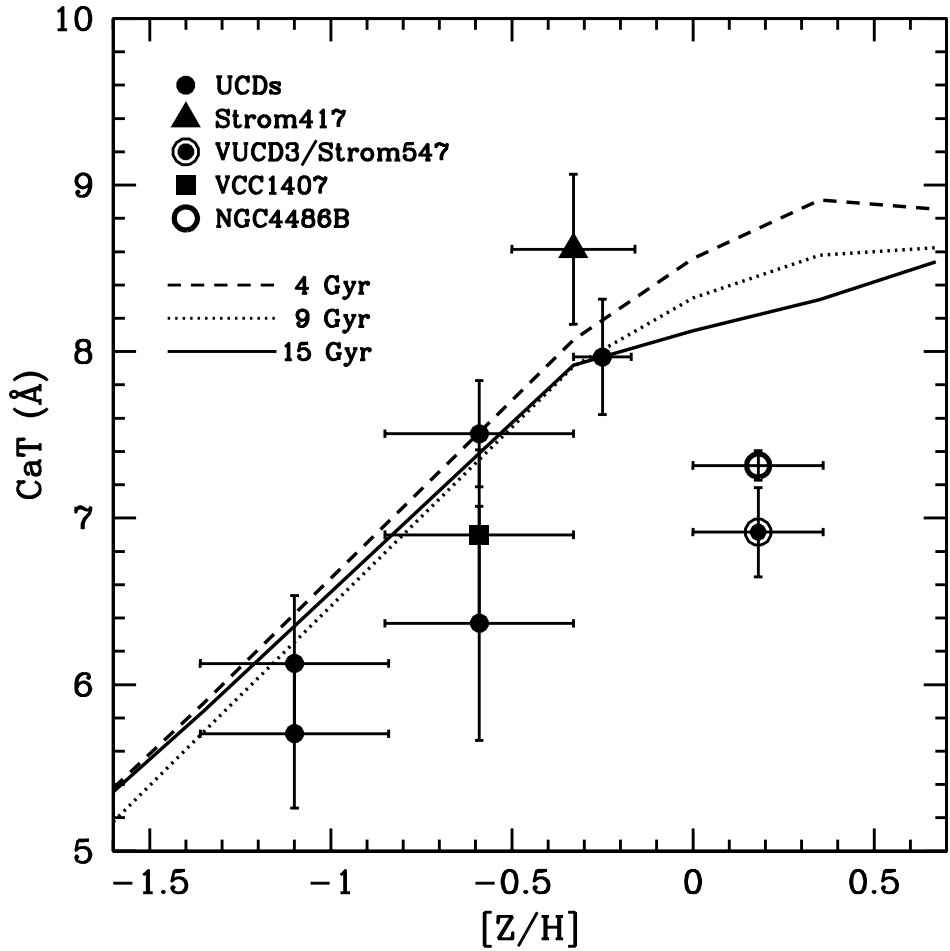}
\caption{Comparison of our data, metallicity vs. CaT index, with 
Maraston (2005) SSP model predictions (Salpeter IMF). 
The SSP models are plotted for ages 4, 9 and 15 Gyr.}
\end{figure}






\clearpage

\begin{deluxetable}{ccccc}
\tabletypesize{\scriptsize}
\tablecaption{Spectroscopy targets.}
\tablewidth{0pt}
\tablehead{
\colhead{Name} & 
\colhead{Object type} & 
\colhead{R.A.(J2000)} & 
\colhead{Dec.(J2000)} & 
\colhead{Exposure times} \\
\colhead{} & 
\colhead{} & 
\colhead{(h$\,$:$\,$m$\,$:$\,$s)} & 
\colhead{($^\circ\,$:$\,'\,$:$\,''$)} & 
\colhead{(s)}
}
\startdata
Strom417 & M87 GC & 12:31:01.29 & +12:19:25.6 & $3\times1800$ \\ 
VUCD1 & UCD & 12:30:07.61 & +12:36:31.1 & $3\times1800$ \\
VUCD3/Strom547 & UCD / M87 GC & 12:30:57.40 & +12:25:44.8 & $3\times1800$ \\
VUCD4 & UCD & 12:31:04.51 & +11:56:36.8 & $3\times1800$ \\
VUCD5 & UCD & 12:31:11.90 & +12:41:01.2 & $3\times1800$ \\
VUCD6 & UCD & 12:31:28.41 & +12:25:03.3 & $3\times1800$ \\
VUCD7 & UCD & 12:31:52.93 & +12:15:59.5 & $2\times1800$ \\
VCC1407 & dE,N & 12:32:02.70 & +11:53:25.0 & $1\times1800+2\times1200$ \\ 
NGC4486B & E & 12:30:31.92 & +12:29:27.4 & $1\times721$ 
\enddata
\end{deluxetable}


\begin{deluxetable}{rclcc}
\tabletypesize{\scriptsize}
\tablecaption{Stellar templates.}
\tablewidth{0pt}
\tablehead{
\colhead{} & 
\colhead{Template} & 
\colhead{SpType} & 
\colhead{[Fe/H]} & 
\colhead{Ref.} 
}
\startdata
 1 & HD040460 & K1III & -0.50 & 2 \\
 2 & HD048433 & K1III & -0.26 & 1 \\
 3 & HD137704 & K4III & -0.43 & 1 \\
 4 & HD139195 & K0III & -0.17 & 1 \\
 5 & HD139669 & K5III & -0.13 & 5 \\
 6 & HD141680 & G8III & -0.28 & 1 \\
 7 & HD142574 & M0III &  ... & ... \\
 8 & HD143107 & K2III & -0.32 & 1 \\
 9 & HD145148 & K1.5IV & ... & ... \\
10 & HD145675 & K0V & 0.31 & 3 \\ 
11 & HD147677 & K0III & -0.08 & 1 \\
12 & HD148513 & K4III & 0.04/-0.31 & 1/4 \\
13 & HD149161 & K4III & -0.23 & 1 
\enddata
\tablerefs{
(1) McWilliam 1990; (2) Cottrell \& Sneden 1986; (3) Peterson 1978; 
(4) Luck \& Challener 1995; (5) Valdes et al. 2004.}
\end{deluxetable}


\begin{deluxetable}{cccccc}
\tabletypesize{\scriptsize}
\tablecaption{Heliocentric line-of-sight velocities and velocity dispersions obtained with the wavelength 
range $8400-8750\mbox{\AA}$ including CaT absorption lines.}
\tablewidth{0pt}
\tablehead{
\colhead{} & 
\multicolumn{3}{c}{Direct Fitting} & 
\multicolumn{2}{c}{Cross Correlation} \\ \\
\colhead{Object} & 
\colhead{v$_{helio}$} & 
\colhead{$\sigma$} & 
\colhead{Templates} & 
\colhead{v$_{helio}$} & 
\colhead{$\sigma$} \\
\colhead{} & 
\colhead{$(km~s^{-1})$} & 
\colhead{$(km~s^{-1})$} & 
\colhead{} & 
\colhead{$(km~s^{-1})$} & 
\colhead{$(km~s^{-1})$} 
}
\startdata
VUCD1 & 1227.8 $\pm$ 1.7 & $\,$~33.8  $\pm$ 1.7 & 6 & 1225.4 $\pm$ 3.7 & 34.0 $\pm$ 1.6 \\
VUCD3/Strom547 & $\,$~710.6 $\pm$ 3.5 & $\,$~37.7 $\pm$ 1.4 & 2,8 & $\,$~711.4 $\pm$ 3.4 & 37.8 $\pm$ 1.6 \\
VUCD4 & $\,$~919.7 $\pm$ 1.7 & $\,$~23.9 $\pm$ 2.2 & 6 & $\,$~916.5 $\pm$ 4.2 & 21.3 $\pm$ 2.2 \\
VUCD5 & 1293.1 $\pm$ 1.7 & $\,$~27.4 $\pm$ 1.7 & 6 & 1290.3 $\pm$ 3.0 & 24.9 $\pm$ 2.0 \\ 
VUCD6 & 2105.3 $\pm$ 1.7 & $\,$~24.6 $\pm$ 1.8 & 6 & 2101.7 $\pm$ 3.9 & 22.2 $\pm$ 2.2 \\
VUCD7 & $\,$~988.3 $\pm$ 2.7 & $\,$~36.7 $\pm$ 3.7 & 6 & $\,$~985.7 $\pm$ 5.0 & 35.6 $\pm$ 1.6 \\
Strom417 & 1863.5 $\pm$ 1.6 & $\,$~26.4 $\pm$ 2.1 & 6 & 1860.6 $\pm$ 3.0 & 25.8 $\pm$ 2.0 \\ 
VCC1407 & 1018.9 $\pm$ 3.2 & $\,$~29.3 $\pm$ 2.5 & 1,2,3,4,6 & 1019.4 $\pm$ 5.7 & 26.6 $\pm$ 2.1 \\
NGC4486B & 1558.4 $\pm$ 4.2 & 211.3 $\pm$ 4.8 & 2,3,7,8,11,12 & $\,$~1546.9 $\pm$ 32.8 & 198.1 $\pm$ 15.5 
\enddata
\tablecomments{In direct-fitting method, the best fitting 
template stars were used to determine velocities and velocity dispersions. The best fitting templates 
(as numbered in Table~2) are given in the ``Templates'' column. 
$\sigma$ and v$_{helio}$ are the mean values using all observations of the best fitting template(s). 
In cross-correlation method, $\sigma$ and v$_{helio}$ are the mean values using all stellar templates.}
\end{deluxetable}


\begin{deluxetable}{cccccc}
\tabletypesize{\scriptsize}
\tablecaption{Heliocentric line-of-sight velocities and velocity dispersions obtained with the wavelength 
range $5100-5250\mbox{\AA}$ including Mgb absorption lines.}
\tablewidth{0pt}
\tablehead{
\colhead{} & 
\multicolumn{3}{c}{Direct Fitting} & 
\multicolumn{2}{c}{Cross Correlation} \\ \\
\colhead{Object} & 
\colhead{v$_{helio}$} & 
\colhead{$\sigma$} & 
\colhead{Templates} & 
\colhead{v$_{helio}$} & 
\colhead{$\sigma$} \\
\colhead{} & 
\colhead{$(km~s^{-1})$} & 
\colhead{$(km~s^{-1})$} & 
\colhead{} & 
\colhead{$(km~s^{-1})$} & 
\colhead{$(km~s^{-1})$} 
}
\startdata
VUCD1 & 1219.5 $\pm$ 1.3 & $\,$~34.4 $\pm$ 1.6 & 2 & 1225.1 $\pm$ 9.3 & 32.8 $\pm$ 6.1 \\
VUCD3/Strom547 & $\,$~716.4 $\pm$ 1.2 & $\,$~46.5 $\pm$ 1.6 & 9 & $\,\,$~~717.4 $\pm$ 12.7 & 51.4 $\pm$ 4.7 \\
VUCD4 & $\,$~912.5 $\pm$ 1.5 & $\,$~23.9 $\pm$ 1.8 & 2 & $\,$~916.8 $\pm$ 5.8 & 15.7 $\pm$ 8.3 \\
VUCD5 & 1285.8 $\pm$ 1.0 & $\,$~28.9 $\pm$ 1.5 & 2 & 1291.8 $\pm$ 6.3 & 26.0 $\pm$ 6.8 \\ 
VUCD6 & 2096.7 $\pm$ 1.4 & $\,$~25.7 $\pm$ 1.8 & 2 & 2103.1 $\pm$ 6.8 & 20.4 $\pm$ 7.6 \\
VUCD7 & $\,$~980.7 $\pm$ 3.5 & $\,$~39.2 $\pm$ 4.4 & 2 & $\,$~986.6 $\pm$ 7.7 & 33.3 $\pm$ 6.0 \\
Strom417 & 1857.1 $\pm$ 1.4 & $\,$~28.8 $\pm$ 1.6 & 2 & 1863.4 $\pm$ 6.7 & 25.1 $\pm$ 6.9 \\ 
VCC1407 & 1018.8 $\pm$ 3.3 & $\,$~31.4 $\pm$ 2.7 & 1,2,6,11 & 1020.2 $\pm$ 7.5 & 28.5 $\pm$ 6.5 \\
NGC4486B & 1556.3 $\pm$ 8.7 & 199.1 $\pm$ 5.0 & 2,6,8,9 & $\,$~1558.7 $\pm$ 52.1 & 230.4 $\pm$ 26.3 
\enddata
\tablecomments{In direct-fitting method, the best fitting 
template stars were used to determine velocities and velocity dispersions. The best fitting templates 
(as numbered in Table~2) are given in the ``Templates'' column. 
$\sigma$ and v$_{helio}$ are the mean values using all observations of the best fitting template(s). 
In cross-correlation method, $\sigma$ and v$_{helio}$ are the mean values using all stellar templates.}
\end{deluxetable}


\begin{deluxetable}{ccc}
\tabletypesize{\scriptsize}
\tablecaption{Adopted radial velocity and velocity dispersion of the Virgo cluster objects.}
\tablewidth{0pt}
\tablehead{
\colhead{Object} & 
\colhead{v$_{helio}$} & 
\colhead{$\sigma$} \\
\colhead{} & 
\colhead{$(km~s^{-1})$} & 
\colhead{$(km~s^{-1})$} 
}
\startdata
Strom417 & 1860.3 $\pm$ 1.5 & $\,$~27.6 $\pm$ 1.9 \\ 
VUCD1 & 1223.7 $\pm$ 1.5 & $\,$~34.1 $\pm$ 1.7 \\
VUCD3/Strom547 & $\,$~713.5 $\pm$ 2.4 & $\,$~42.1 $\pm$ 1.5 \\
VUCD4 & $\,$~916.1 $\pm$ 1.6 & $\,$~23.9 $\pm$ 2.0 \\
VUCD5 & 1289.5 $\pm$ 1.4 & $\,$~28.2 $\pm$ 1.6 \\
VUCD6 & 2101.0 $\pm$ 1.6 & $\,$~25.2 $\pm$ 1.8 \\
VUCD7 & $\,$~984.5 $\pm$ 3.1 & $\,$~38.0 $\pm$ 4.1\\
VCC1407 & 1018.9 $\pm$ 3.3 & $\,$~30.4 $\pm$ 2.6 \\ 
NGC4486B & 1557.4 $\pm$ 6.5 & 205.2 $\pm$ 4.9 
\enddata
\end{deluxetable}


\begin{deluxetable}{cccc}
\tabletypesize{\scriptsize}
\tablecaption{Virgo UCD photometry.}
\tablewidth{0pt}
\tablehead{
\colhead{Name} & 
\colhead{$m_V$} & 
\colhead{$M_V$} & 
\colhead{$V - I$} \\
\colhead{} & 
\colhead{(mag)} & 
\colhead{(mag)} & 
\colhead{(mag)} 
}
\startdata
VUCD1 &  18.66 & -12.26 &  0.96 \\
VUCD3/Strom547  & 18.34 & -12.58 & 1.27 \\
VUCD4 &  18.62 & -12.30 &  0.99 \\
VUCD5 &  18.60 & -12.32 &  1.11 \\
VUCD6 &  18.82 & -12.10 &  1.02 \\
VUCD7 &  17.48 & -13.44 &  1.13 
\enddata
\tablecomments{The $V$ band apparent magnitude, $m_V$, is determined as 
described in Section 3 and is corrected for foreground dust extinction (Schlegel et al. 1998).
The absolute magnitude, $M_V$, is computed assuming a Virgo Cluster distance modulus of 30.92~mag 
(Freedman et al. 2001). The $V - I$ color is reddening corrected.}
\end{deluxetable}


\begin{deluxetable}{lcccc}
\tabletypesize{\scriptsize}
\tablecaption{Fornax UCDs and a Fornax dE,N (FCC303): photometry.}
\tablewidth{0pt}
\tablehead{
\colhead{Object} & 
\colhead{R.A.(J2000)} & 
\colhead{Dec.(J2000)} & 
\colhead{$m_V$} & 
\colhead{$M_V$} \\
\colhead{} & 
\colhead{(h$\,$:$\,$m$\,$:$\,$s)} & 
\colhead{($^\circ\,$:$\,'\,$:$\,''$)} & 
\colhead{(mag)} & 
\colhead{(mag)} 
}
\startdata
UCD1 & 3:37:03.30 & -35:38:04.6 & 19.20 & -12.19 \\
UCD2 & 3:38:06.33 & -35:28:58.8 & 19.12 & -12.27 \\
UCD3 & 3:38:54.10 & -35:33:33.6 & 17.82 & -13.57 \\
UCD4 & 3:39:35.95 & -35:28:24.5 & 18.94 & -12.45 \\
UCD5 & 3:39:52.58 & -35:04:24.1 & 19.40 & -11.99 \\
FCC303 & 3:45:14.08 & -36:56:12.4 & 15.90 & -15.49 
\enddata
\tablecomments{ The $V$ band apparent magnitude, $m_V$, is determined as 
described in Section 3 and is corrected for foreground dust extinction (Schlegel et al. 1998).
The absolute magnitude, $M_V$, is computed assuming a Fornax Cluster distance modulus of 31.39~mag 
(Freedman et al. 2001).}
\end{deluxetable}


\begin{deluxetable}{lccccccc}
\tabletypesize{\scriptsize}
\tablecaption{Structural parameters for Virgo UCDs from ACS/HRC photometry.}
\tablewidth{0pt}
\tablehead{
\colhead{} & 
\colhead{VUCD1} & 
\colhead{VUCD3} & 
\colhead{VUCD4} & 
\colhead{VUCD5} & 
\colhead{VUCD6} & 
\colhead{VUCD7 core} & 
\colhead{VUCD7 halo\tablenotemark{a}}
}
\startdata
$R_{e,obs}$        & $\,$~15.4 & $\,$~28.1 & $\,$~29.4 & $\,$~22.5 & $\,$~22.1 & ... & ...  \\ \\
{\bf NUKER}        & & & & & & &   \\  
$\chi_{\nu}^2$ (F606W)   & $\,$~0.39 & $\,$~0.41 & $\,$~0.35 & $\,$~0.38 & $\,$~0.24 & ... & ...   \\
$\chi_{\nu}^2$ (F814W)   & $\,$~0.38 & $\,$~0.43 & $\,$~0.34 & $\,$~0.36 & $\,$~0.27 & ... & ...   \\
$R_e$              & $\,$~13.3 & ... & ... & $\,$~19.2 & ... & ... & ...  \\
$m_{V,{\rm tot}}$  & 18.49 & ... & ... & 18.55 & ... & ... & ...  \\ 
$R_b$              & $\,\,$~~6.3 & $\,$~73.3 & $\,\,$~~9.1 & $\,$~19.3 & $\,\,$~~5.2 & ... & ...  \\
$\mu_V(R_b)$       & 16.42 & 22.00 & 17.24 & 18.73 & 16.45 & ... & ...  \\
$\alpha$           & $\,$~2.52 & $\,$~1.95 & $\,$~3.92 & $\,$~0.99 & 19.48 & ... & ...  \\
$\beta$            & $\,$~2.69 & $\,$~3.17 & $\,$~2.34 & $\,$~4.02 & $\,$~2.24 & ... & ...  \\
$\gamma$           & $\,$~0.23 & $\,$~1.53 & $\,$~0.33 & $\,$~0.00 & $\,$~0.50 & ... & ...  \\
$\epsilon$         & $\,$~0.06 & $\,$~0.15 & $\,$~0.15 & $\,$~0.01 & $\,$~0.04 & ... & ...  \\ 
{\bf SERSIC}      & & & & & & &   \\
$\chi_{\nu}^2$ (F606W)    & $\,$~0.48 & $\,$~0.42 & $\,$~0.40 & $\,$~0.39 & $\,$~0.29 & $\,$~0.37 & $\,$~0.37/0.37   \\
$\chi_{\nu}^2$ (F814W)    & $\,$~0.44 & $\,$~0.45 & $\,$~0.38 & $\,$~0.38 & $\,$~0.31 & $\,$~0.37 & $\,$~0.36/0.37   \\
$R_e$              & $\,$~11.1 & $\,$~64.3 & $\,$~19.3 & $\,$~18.1 & $\,$~12.9 & $\,\,$~~9.2 & $\,$~214.1/223.4  \\
$m_{V,{\rm tot}}$  & 18.75 & 17.69 & 18.85 & 18.64 & 19.05 & 18.68 & $\,$~18.01/17.85  \\ 
n                  & $\,\,$~~2.2 & $\,$~10.9 & $\,\,$~~2.1 & $\,\,$~~1.9 & $\,\,$~~3.1 & $\,\,$~~2.2 & $\,$~1.4/2.1  \\
$\epsilon$         & $\,$~0.07 & $\,$~0.16 & $\,$~0.16 & $\,$~0.01 & $\,$~0.04 & $\,$~0.12 & $\,$~0.05/0.05  \\ 
{\bf KING, $\alpha = 2$}        & & & & & & &   \\
$\chi_{\nu}^2$ (F606W)    & $\,$~0.45 & $\,$~0.55 & $\,$~0.37 & $\,$~0.37 & $\,$~0.27 & $\,$~0.37 & ...   \\
$\chi_{\nu}^2$ (F814W)    & $\,$~0.42 & ... & $\,$~0.36 & $\,$~0.37 & $\,$~0.29 & $\,$~0.36 & ...   \\
$R_e$              & $\,$~11.2 & ... & $\,$~21.8 & $\,$~17.8 & $\,$~15.2 & $\,$~10.4 & ...  \\
$m_{V,{\rm tot}}$  & 18.66 & ... & 18.54 & 18.59 & 18.93 & 18.42 & ...  \\ 
$R_c$              & $\,\,$~~3.6 & $\,\,$~~1.9 & $\,\,$~~5.8 & $\,\,$~~6.6 & $\,\,$~~2.7 & $\,\,$~~3.1 & ...  \\
$R_t$              & 124.0 & $\,\,$~~$\infty$ & 302.7 & 172.5 & 355.6 & 130.2 & ...  \\
c                  & $\,$~1.54 & ... & $\,$~1.71 & $\,$~1.42 & $\,$~2.12 & $\,$~1.62 & ...  \\
$\mu_{0,V}$        & 14.91 & 13.79 & 15.96 & 15.98 & 14.98 & 14.38 & ...  \\
$\epsilon$         & $\,$~0.06 & $\,$~0.17 & $\,$~0.15 & $\,$~0.01 & $\,$~0.04 & $\,$~0.11 & ...  \\ 
\multicolumn{2}{l}{\bf KING with variable $\alpha$:}  & & & & & &  \\
$\chi_{\nu}^2$ (F606W)    & $\,$~0.41 & $\,$~0.53 & $\,$~0.36 & $\,$~0.37 & $\,$~0.25 & ... & ...   \\
$\chi_{\nu}^2$ (F814W)    & $\,$~0.40 & ... & $\,$~0.35 & $\,$~0.37 & $\,$~0.28 & ... & ...   \\
$\alpha$           & $\,$~3.74 & $\,$~0.63 & $\,$~3.95 & $\,$~2.24 & $\,$~4.45 & ... & ... \\
$R_e$              & $\,$~11.3 & $\,$~18.7 & $\,$~22.0 & $\,$~17.9 & $\,$~14.8 & ... & ...  \\
$m_{V,{\rm tot}}$  & 18.63 & 18.18 & 18.52 & 18.60 & 18.84 & ... & ...  \\ 
$R_c$              & $\,\,$~~4.3 & $\,\,$~~1.8 & $\,\,$~~6.7 & $\,\,$~~6.7 & $\,\,$~~3.2 & ... & ...  \\
$R_t$              & 360.0 & 247.5 & 1217.4 & 200.5 & 2352.5 & ... & ...  \\
c                  & $\,$~1.92 & $\,$~2.14 & $\,$~2.26 & $\,$~1.48 & $\,$~2.87 & ... & ...  \\
$\mu_{0,V}$        & 14.68 & 13.78 & 15.76 & 15.97 & 14.81 & ... & ...  \\
$\epsilon$         & $\,$~0.06 & $\,$~0.17 & $\,$~0.15 & $\,$~0.01 & $\,$~0.04 & ... & ...  \\ \\
Best model    & $\,$~N & $\,$~N & $\,$~N & $\,$~N & $\,$~N & $\,$~K & $\,$~S  
\enddata
\tablenotetext{a}{The first number is for King+Sersic model, the second number is for Sersic+Sersic model.}
\tablecomments{All the parameters 
are the mean of the two passbands, $V$ and $I$, except King models for VUCD3 (see explanation in the text).
 Units: $R_{e,obs}$, $R_e$, $R_b$, $R_c$ and $R_t$ are in pc;  
$\mu_V(R_b)$ and $\mu_{0,V}$  are in mag arcsec$^{-2}$; $m_{V,tot}$ is in mag.
$\mu_V(R_b)$, $\mu_{0,V}$ and $m_{V,tot}$ are corrected for the extinction in our Galaxy.}
\end{deluxetable}


\begin{deluxetable}{lcccccccccc}
\tabletypesize{\scriptsize}
\tablecaption{Structural parameters for Fornax UCDs and a dE,N (FCC303) from STIS photometry.}
\tablewidth{0pt}
\tablehead{
\colhead{} & 
\colhead{UCD1} & 
\colhead{UCD2} & 
\colhead{UCD3} & 
\colhead{UCD3} & 
\colhead{UCD3} & 
\colhead{UCD4} & 
\colhead{UCD5} & 
\colhead{UCD5} & 
\colhead{FCC303} & 
\colhead{FCC303} \\ 
\colhead{} & 
\colhead{} & 
\colhead{} & 
\colhead{} & 
\colhead{core} & 
\colhead{halo\tablenotemark{a}} & 
\colhead{} & 
\colhead{core} & 
\colhead{halo\tablenotemark{a}} & 
\colhead{core} & 
\colhead{halo\tablenotemark{a}} 
}
\startdata
$R_{e,obs}$        &  $\,$~33.2 &  $\,$~29.5 &  $\,$~80.7 & ... & ... &  $\,$~31.3 & ... & ... & ... & ... \\ \\
{\bf NUKER}        & & & & & & & & & &  \\ 
$\chi_{\nu}^2$     & $\,$~0.61 & $\,$~0.64 & $\,$~0.51 & ... & ... & $\,$~0.50 & ... & ... & ... & ...   \\ 
$R_e$              & ... & ... & ... & ... & ... & ... & ... & ... & ... & ... \\
$m_{V,{\rm tot}}$  & ... & ... & ... & ... & ... & ... & ... & ... & ... & ... \\ 
$R_b$              & $\,$~36.7 & $\,\,$~~4.3 & 318.7 & ... & ... & $\,\,$~~7.3 & ... & ... & ... & ... \\
$\mu_V(R_b)$       & 20.67 & 16.53 & 24.38 & ... & ... & 16.92 & ... & ... & ... & ... \\
$\alpha$           & $\,$~0.34 & $\,$~0.73 & $\,$~1.90 & ... & ... & 20.00 & ... & ... & ... & ... \\
$\beta$            & $\,$~3.03 & $\,$~2.38 & $\,$~7.04 & ... & ... & $\,$~2.13 & ... & ... & ... & ... \\
$\gamma$           & $\,$~0.91 & $\,$~0.58 & $\,$~1.10 & ... & ... & $\,$~0.88 & ... & ... & ... & ... \\
$\epsilon$         & $\,$~0.18 & $\,$~0.01 & $\,$~0.03 & ... & ... & $\,$~0.05 & ... & ... & ... & ... \\ 
{\bf SERSIC}       & & & & & & & & & &  \\
$\chi_{\nu}^2$     & $\,$~0.60 & $\,$~0.66 & ... & $\,$~0.47 & $\,$~0.47/0.47 & $\,$~0.69 & $\,$~0.47 & $\,$~0.47/0.47 & $\,$~1.03 & $\,$~1.03/1.03   \\
$R_e$              & $\,$~36.9 & $\,$~26.6 & ... & $\,\,$~~8.6 & $\,$~106.6/103.2 & $\,$~24.1 & $\,\,$~~6.0 & $\,$~134.5/135.5 & $\,$~25.2 & $\,$~696.3/692.8 \\
$m_{V,{\rm tot}}$  & 19.01 & 19.03 & ... & 20.29 & $\,$~17.98/17.94 & 18.97 & 20.45 & $\,$~19.54/19.58 & 18.76 & $\,$~15.95/15.95 \\ 
n                  & $\,\,$~~9.9 & $\,\,$~~6.8 & ... & $\,\,$~~1.7 & $\,$~1.3/1.5 & $\,\,$~~5.5 & $\,\,$~~1.1 & $\,$~6.9/6.3 & $\,$~10.7 & $\,$~0.6/0.6 \\
$\epsilon$         & $\,$~0.19 & $\,$~0.01 & ... & $\,$~0.02 & $\,$~0.03/0.03 & $\,$~0.05 & $\,$~0.24 & $\,$~0.16/0.15 & $\,$~0.03 & $\,$~0.10/0.10 \\ 
{\bf KING, $\alpha = 2$}        & & & & & & & & & &  \\
$\chi_{\nu}^2$     & $\,$~0.67 & $\,$~0.63 & ... & $\,$~0.47 & ... & $\,$~0.52 & $\,$~0.47 & ... & $\,$~1.03 & ...   \\
$R_e$              & $\,$~48.9 & $\,$~28.3 & ... & $\,$~10.9 & ... & $\,$~26.0 & $\,\,$~~6.0 & ... & $\,$~25.9 & ... \\
$m_{V,{\rm tot}}$  & 18.71 & 18.98 & ... & 20.00 & ... & 18.88 & 20.19 & ... & 18.78 & ... \\ 
$R_c$              & $\,\,$~~1.8 & $\,\,$~~2.3 & ... & $\,\,$~~3.6 & ... & $\,\,$~~2.8 & $\,\,$~~4.0 & ... & $\,\,$~~1.2 & ... \\
$R_t$              & 5761.2 & 1457.5 & ... & 119.2 & ... & 987.6 & $\,$~30.7 & ... & 2403.9 & ... \\
c                  & $\,$~3.51 & $\,$~2.80 & ... & $\,$~1.52 & ... & $\,$~2.55 & $\,$~0.89 & ... & $\,$~3.30 & ... \\
$\mu_{0,V}$        & 14.21 & 14.70 & ... & 15.74 & ... & 14.89 & 15.13 & ... & 13.32 & ... \\
$\epsilon$         & $\,$~0.18 & $\,$~0.01 & ... & $\,$~0.03 & ... & $\,$~0.05 & $\,$~0.24 & ... & $\,$~0.03 & ... \\ 
\multicolumn{2}{l}{\bf KING with variable $\alpha$:}  & & & & & & & & &  \\
$\chi_{\nu}^2$     & $\,$~0.64 & $\,$~0.62 & ... & ... & ... & $\,$~0.51 & ... & ... & ... & ...   \\
$\alpha$           & $\,$~0.74 & $\,$~1.23 & ... & ... & ... & $\,$~3.32 & ... & ... & ... & ...  \\
$R_e$              & $\,$~22.4 & $\,$~23.1 & ... & ... & ... & $\,$~29.5 & ... & ... & ... & ...  \\
$m_{V,{\rm tot}}$  & 19.00 & 19.10 & ... & ... & ... & 18.82 & ... & ... & ... & ... \\ 
$R_c$              & $\,\,$~~1.8 & $\,\,$~~2.2 & ... & ... & ... & $\,\,$~~3.0 & ... & ... & ... & ... \\
$R_t$              & 378.0 & 487.1 & ... & ... & ... & 4501.2 & ... & ... & ... & ... \\
c                  & $\,$~2.32 & $\,$~2.35 & ... & ... & ... & $\,$~3.18 & ... & ... & ... & ... \\
$\mu_{0,V}$        & 14.21 & 14.67 & ... & ... & ... & 14.94 & ... & ... & ... & ... \\
$\epsilon$         & $\,$~0.18 & $\,$~0.01 & ... & ... & ... & $\,$~0.05 & ... & ... & ... & ... \\ \\
Best model         & $\,$~S & gen.~K & K+S/S+S & $\,$~K & $\,$~S & $\,$~N & $\,$~K/S & $\,$~S & $\,$~K/S & $\,$~S 
\enddata
\tablenotetext{a}{The first number is for King+Sersic model, the second number is for Sersic+Sersic model.}
\tablecomments{Units: $R_{e,obs}$, $R_e$, $R_b$, $R_c$ and $R_t$ are in pc;  
$\mu_V(R_b)$ and $\mu_{0,V}$  are in mag arcsec$^{-2}$; $m_{V,tot}$ is in mag.
$\mu_V(R_b)$, $\mu_{0,V}$ and $m_{V,tot}$ are corrected for the extinction in our Galaxy.}
\end{deluxetable}


\begin{deluxetable}{lcccc}
\tabletypesize{\scriptsize}
\tablecaption{Adopted photometric parameters for Virgo and Fornax UCDs and FCC303 galaxy.}
\tablewidth{0pt}
\tablehead{
\colhead{} & 
\colhead{$R_{\rm eff}$} & 
\colhead{$m_V$} & 
\colhead{$\left<\mu_V\right>_{\rm eff}$} & 
\colhead{$\epsilon$} \\ 
\colhead{} & 
\colhead{(pc)} & 
\colhead{(mag)} & 
\colhead{(mag arcsec$^{-2}$)} & 
\colhead{} 
}
\startdata
{\bf Virgo:}        & & & & \\  
VUCD1                    & $\,$~11.3  & 18.66 & 16.58 & 0.06 \\
VUCD3                    & $\,$~18.7  & 18.34 & 17.34 & 0.15 \\
VUCD4                    & $\,$~22.0  & 18.62 & 17.98 & 0.15 \\
VUCD5                    & $\,$~17.9  & 18.60 & 17.51 & 0.01 \\
VUCD6                    & $\,$~14.8  & 18.82 & 17.32 & 0.04 \\
VUCD7                    & $\,$~96.8  & 17.48 & 20.06 & ... \\
VUCD7 core               & $\,$~10.4  & 18.42 & 16.15 & 0.11 \\
VUCD7 halo               & 214.1      & 18.01 & 22.31 & 0.05 \\ \\
{\bf Fornax:}       & & & & \\
UCD1                     & $\,$~22.4  & 19.20 & 18.13 & 0.19 \\
UCD2                     & $\,$~23.1  & 19.12 & 18.12 & 0.01 \\ 
UCD3                     & $\,$~89.7  & 17.82 & 19.76 & ... \\
UCD3 core                & $\,$~10.9  & 20.00 & 17.36 & 0.03 \\
UCD3 halo                & 106.6      & 17.98 & 20.30 & 0.03 \\
UCD4                     & $\,$~29.5  & 18.94 & 18.47 & 0.05 \\
UCD5                     & $\,$~31.2  & 19.40 & 19.05 & ... \\
UCD5 core             & $\,\,$~~6.0   & 20.19 & 16.25 & 0.24 \\
UCD5 halo                & 134.5      & 19.54 & 22.37 & 0.16 \\
FCC303                   & 660.0      & 15.90 & 22.18 & ... \\
FCC303 core              & $\,$~25.9  & 18.78 & 18.03 & 0.03 \\ 
FCC303 halo              & 696.3      & 15.95 & 22.35 & 0.10 
\enddata
\end{deluxetable}


\begin{deluxetable}{lcccc}
\tabletypesize{\scriptsize}
\tablecaption{Dynamical modeling results for different light profile representations.
}
\tablewidth{0pt}
\tablehead{
\colhead{} & 
\colhead{$\sigma_0$} & 
\colhead{$\sigma$} & 
\colhead{$M$} & 
\colhead{$M/L_V$} \\
\colhead{} & 
\colhead{$(km~s^{-1})$} & 
\colhead{$(km~s^{-1})$} & 
\colhead{($10^7M_\odot$)} & 
\colhead{($M_\odot/L_\odot$)}
}
\startdata
\multicolumn{2}{l}{\bf Nuker:} & & & \\
VUCD1 & $40.0\pm1.2$ & $30.7\pm2.6$ & $3.2\pm0.5$ & $4.5\pm0.8$ \\
VUCD3 & $81.3\pm15.$ & $33.7\pm10.4$ & $1.8\pm0.9$ & $2.0\pm1.0$ \\
VUCD4 & $27.3\pm1.5$ & $19.7\pm3.0$ & $2.9\pm0.7$ & $4.0\pm1.1$ \\
VUCD5 & $31.5\pm1.5$ & $26.1\pm2.5$ & $3.0\pm0.5$ & $4.1\pm0.8$ \\
VUCD6 & $28.3\pm1.0$ & $20.6\pm3.6$ & $2.2\pm0.6$ & $3.7\pm1.1$ \\[1mm]
\multicolumn{2}{l}{\bf Sersic:} & & & \\
VUCD1 & $40.1\pm1.6$ & $32.7\pm2.2$ & $2.6\pm0.3$ & $3.8\pm0.6$ \\
VUCD3 & $61.0\pm4.5$ & $31.1\pm9.0$ & $5.7\pm2.6$ & $6.1\pm2.8$ \\
VUCD4 & $26.8\pm1.4$ & $22.3\pm2.5$ & $2.1\pm0.5$ & $3.0\pm0.7$ \\
VUCD5 & $31.5\pm1.6$ & $26.5\pm1.9$ & $2.9\pm0.4$ & $3.9\pm0.6$ \\
VUCD6 & $30.2\pm0.8$ & $23.4\pm3.1$ & $1.5\pm0.4$ & $2.5\pm0.7$ \\[1mm]
\multicolumn{2}{l}{\bf King, $\alpha=2$:} & & & \\
VUCD1 & $39.8\pm0.8$ & $32.6\pm2.4$ & $2.7\pm0.4$ & $3.9\pm 0.6$ \\
VUCD4 & $27.7\pm1.1$ & $21.8\pm2.8$ & $2.4\pm0.6$ & $3.3\pm0.8$ \\
VUCD5 & $31.6\pm0.7$ & $26.4\pm2.0$ & $2.9\pm0.4$ & $3.9\pm0.6$ \\
VUCD6 & $30.5\pm0.6$ & $22.7\pm5.3$ & $1.8\pm0.7$ & $2.9\pm1.3$ \\
Strom417 & $31.7\pm1.4$ & $26.4\pm2.7$ & $2.7\pm0.5$ & $6.6\pm1.5$ \\[1mm]
\multicolumn{2}{l}{\bf King with variable $\alpha$:} & & & \\
VUCD1 & $39.3\pm2.0$ & $32.2\pm2.4$ & $2.8\pm0.5$ & $4.0\pm0.7$ \\
VUCD3 & $52.2\pm2.5$ & $35.8\pm1.5$ & $5.0\pm0.7$ & $5.4\pm0.9$ \\
VUCD4 & $26.9\pm2.3$ & $21.3\pm2.0$ & $2.4\pm0.6$ & $3.4\pm0.9$ \\
VUCD5 & $32.5\pm2.3$ & $26.4\pm1.6$ & $2.9\pm0.4$ & $3.9\pm0.6$ \\
VUCD6 & $29.6\pm2.2$ & $22.3\pm1.8$ & $1.8\pm0.5$ & $2.9\pm0.9$ \\[1mm]
\multicolumn{2}{l}{\bf King core + Sersic halo:} & & & \\
VUCD7 & $45.1\pm1.5$ & $27.2\pm4.6$ & $8.8\pm2.1$ & $4.3\pm1.1$ 
\enddata
\tablecomments{$\sigma_0$ -- central velocity dispersion, 
$\sigma$ -- global velocity dispersion, $M$ -- mass, $M/L_V$ -- mass-to-light ratio.}
\end{deluxetable}


\begin{deluxetable}{lcccc}
\tabletypesize{\scriptsize}
\tablecaption{Adopted velocity dispersions (central and global), masses and
mass-to-light ratios for Virgo UCDs.}
\tablewidth{0pt}
\tablehead{
\colhead{} & 
\colhead{$\sigma_0$} & 
\colhead{$\sigma$} & 
\colhead{$M$} & 
\colhead{$M/L_V$} \\
\colhead{} & 
\colhead{$(km~s^{-1})$} & 
\colhead{$(km~s^{-1})$} & 
\colhead{($10^7M_\odot$)} & 
\colhead{($M_\odot/L_\odot$)} 
}
\startdata
VUCD1 & $39.3\pm2.0$ & $32.2\pm2.4$ & $2.8\pm0.5$ & $4.0\pm0.7$ \\
VUCD3 & $52.2\pm2.5$ & $35.8\pm1.5$ & $5.0\pm0.7$ & $5.4\pm0.9$ \\
VUCD4 & $26.9\pm2.3$ & $21.3\pm2.0$ & $2.4\pm0.6$ & $3.4\pm0.9$ \\
VUCD5 & $32.5\pm2.3$ & $26.4\pm1.6$ & $2.9\pm0.4$ & $3.9\pm0.6$ \\
VUCD6 & $29.6\pm2.2$ & $22.3\pm1.8$ & $1.8\pm0.5$ & $2.9\pm0.9$ \\
VUCD7 & $45.1\pm1.5$ & $27.2\pm4.6$ & $8.8\pm2.1$ & $4.3\pm1.1$ \\
Strom417 & $31.7\pm1.4$ & $26.4\pm2.7$ & $2.7\pm0.5$ & $6.6\pm1.5$ 
\enddata
\tablecomments{The values are based on the generalized
King models for the one-component fits and King+Sersic models for the two-component fits.}
\end{deluxetable}


\begin{deluxetable}{lcccc}
\tabletypesize{\scriptsize}
\tablecaption{Masses and mass-to-light ratios from King and virial mass estimators.}
\tablewidth{0pt}
\tablehead{
\colhead{} & 
\colhead{$M_K$} & 
\colhead{$M_K/L_V$} & 
\colhead{$M_{\rm vir}$} & 
\colhead{$M_{\rm vir}/L_V$} \\
\colhead{} & 
\colhead{($10^7M_\odot$)} & 
\colhead{($M_\odot/L_\odot$)} & 
\colhead{($10^7M_\odot$)} & 
\colhead{($M_\odot/L_\odot$)}
}
\startdata
VUCD1 & $2.9\pm0.1$ & $4.1\pm0.3$ & $\,$~$2.7\pm0.4$ & $3.8\pm0.7$ \\
VUCD3 & ... & ... & $\,$~$5.4\pm0.7$ & $5.8\pm0.8$ \\
VUCD4 & $2.5\pm0.2$ & $3.5\pm0.3$ & $\,$~$2.3\pm0.5$ & $3.1\pm0.7$ \\
VUCD5 & $2.8\pm0.1$ & $3.7\pm0.2$ & $\,$~$2.8\pm0.4$ & $3.8\pm0.6$ \\
VUCD6 & $2.1\pm0.1$ & $3.5\pm0.3$ & $\,$~$1.7\pm0.4$ & $2.8\pm0.8$ \\
VUCD7 & ... & ... & $16.2\pm 5.7$ & $7.9\pm2.9$ \\
Strom417 & $2.2\pm0.2$ & $5.6\pm0.9$ & $\,$~$2.2\pm0.5$ & $5.4\pm1.3$ 
\enddata
\end{deluxetable}


\begin{deluxetable}{lcccccccc}
\tabletypesize{\scriptsize}
\tablecaption{Mass-to-light ratios and $V - I$ colors for Virgo UCDs from SSP models by Maraston 
(2005) in comparison to the dynamical mass-to-light ratios and observed $V - I$ colors.}
\tablewidth{0pt}
\tablehead{
\colhead{} & 
\colhead{age range} & 
\colhead{met. range} & 
\colhead{$(V-I)_{obs}$} & 
\colhead{$(V-I)_{sp}$} & 
\colhead{$(V-I)_{kr}$} &
\colhead{$(M/L_V)_{dyn}$} & 
\colhead{$(M/L_V)_{sp}$} & 
\colhead{$(M/L_V)_{kr}$} \\
\colhead{} & 
\colhead{(Gyr)} & 
\colhead{(dex)} & 
\colhead{(mag)} & 
\colhead{(mag)} & 
\colhead{(mag)} & 
\colhead{($M_\odot/L_\odot$)} & 
\colhead{($M_\odot/L_\odot$)} & 
\colhead{($M_\odot/L_\odot$)}
}
\startdata
VUCD1 &  $\,$~8...15 & $\,$~$-1.35$...$-0.33$ & 0.96 & 0.91...1.16 & 0.89...1.13 & 
 $4.0\pm0.7$ & $4.5\pm1.9$ & $2.9\pm1.2$ \\
VUCD3 & 12...15 & $\,$~$0.00$...$0.35$   & 1.27 & 1.19...1.30 & 1.16...1.27 & 
 $5.4\pm0.9$ & $8.7\pm2.1$ & $5.6\pm1.4$ \\ 
VUCD4 &  $\,$~8...15 & $\,$~$-1.35$...$-0.33$ & 0.99 & 0.91...1.16 & 0.89...1.13 & 
 $3.4\pm0.9$ & $4.5\pm1.9$ & $2.9\pm1.2$ \\ 
VUCD5 &  $\,$~8...15 & $-0.33$...$0.00$  & 1.11 & 1.07...1.22 & 1.05...1.20 & 
 $3.9\pm0.6$ & $6.1\pm2.3$ & $3.9\pm1.5$ \\ 
VUCD6 &  $\,$~8...15 & $\,$~$-1.35$...$-0.33$ & 1.02 & 0.91...1.16 & 0.89...1.13 & 
 $2.9\pm0.9$ & $4.5\pm1.9$ & $2.9\pm1.2$ \\ 
VUCD7 &  $\,$~8...15 & $\,$~$-1.35$...$-0.33$ & 1.13 & 0.91...1.16 & 0.89...1.13 & 
 $4.3\pm1.1$ & $4.5\pm1.9$ & $2.9\pm1.2$ \\ 
Strom417 & $\,$~4...12 & $-1.35$...$0.00$ & ... & 0.82...1.19 & 0.81...1.16 & 
 $6.6\pm1.5$ & $4.1\pm2.6$ & $2.6\pm1.6$ 
\enddata
\tablecomments{ ``sp'' denotes results for a Salpeter IMF, ``kr'' -- for a Kroupa IMF.}
\end{deluxetable}


\begin{deluxetable}{lcccccc}
\tabletypesize{\scriptsize}
\tablecaption{Lick/IDS indices.}
\tablewidth{0pt}
\tablehead{
\colhead{} & 
\colhead{H$\beta\,(\mbox{\AA})$} & 
\colhead{Mgb$\,(\mbox{\AA})$} & 
\colhead{Fe5270$\,(\mbox{\AA})$} & 
\colhead{Fe5335$\,(\mbox{\AA})$} & 
\colhead{$<$Fe$>$\,$(\mbox{\AA})$} & 
\colhead{$[\rm MgFe]'\,(\mbox{\AA})$} 
}
\startdata
VUCD1        & 2.06 $\pm$ 0.14 & 2.20 $\pm$ 0.13 & 1.56 $\pm$ 0.15 & 1.19 $\pm$ 0.17 & 1.38 $\pm$ 0.16 & 1.79 $\pm$ 0.15 \\
VUCD3/Strom547        & 1.40 $\pm$ 0.14 & 4.99 $\pm$ 0.12 & 2.69 $\pm$ 0.13 & 2.37 $\pm$ 0.15 & 2.53 $\pm$ 0.14 & 3.60 $\pm$ 0.14 \\
VUCD4        & 2.16 $\pm$ 0.18 & 1.42 $\pm$ 0.16 & 1.41 $\pm$ 0.18 & 1.18 $\pm$ 0.21 & 1.30 $\pm$ 0.20 & 1.38 $\pm$ 0.17 \\
VUCD5        & 1.82 $\pm$ 0.17 & 3.30 $\pm$ 0.15 & 2.11 $\pm$ 0.17 & 1.90 $\pm$ 0.19 & 2.01 $\pm$ 0.18 & 2.60 $\pm$ 0.17 \\
VUCD6        & 2.27 $\pm$ 0.16 & 1.58 $\pm$ 0.15 & 1.40 $\pm$ 0.17 & 0.87 $\pm$ 0.19 & 1.14 $\pm$ 0.18 & 1.41 $\pm$ 0.17 \\
VUCD7        & 1.90 $\pm$ 0.32 & 2.45 $\pm$ 0.29 & 1.75 $\pm$ 0.33 & 1.09 $\pm$ 0.38 & 1.42 $\pm$ 0.36 & 1.96 $\pm$ 0.33 \\
Strom417    & 2.18 $\pm$ 0.23 & 2.86 $\pm$ 0.20 & 1.82 $\pm$ 0.23 & 1.47 $\pm$ 0.25 & 1.65 $\pm$ 0.24 & 2.22 $\pm$ 0.23 \\
VCC1407     & 2.08 $\pm$ 0.24 & 2.20 $\pm$ 0.21 & 1.63 $\pm$ 0.24 & 1.53 $\pm$ 0.27 & 1.58 $\pm$ 0.26 & 1.88 $\pm$ 0.24 \\
NGC4486B    & 1.39 $\pm$ 0.04 & 5.24 $\pm$ 0.04 & 2.93 $\pm$ 0.04 & 2.60 $\pm$ 0.05 & 2.77 $\pm$ 0.05 & 3.86 $\pm$ 0.04 
\enddata
\end{deluxetable}


\begin{deluxetable}{lcccccccc}
\tabletypesize{\scriptsize}
\tablecaption{The mean offsets between our instrumental system and the Lick/IDS system.}
\tablewidth{0pt}
\tablehead{
\colhead{Index} & 
\colhead{Lick--ESI} & 
\colhead{rms scatter} & 
\colhead{Lick rms} \\                   
\colhead{(units)} & \colhead{} & \colhead{} & \colhead{per observ.} 
}
\startdata
H$\beta\,(\mbox{\AA})$ & $\,$~0.00 & 0.12 & 0.22 \\
Mgb$\,(\mbox{\AA})$ & -0.13 & 0.05 & 0.23 \\
Fe5270$\,(\mbox{\AA})$ & -0.04 & 0.09 & 0.28 \\
Fe5335$\,(\mbox{\AA})$ & -0.08 & 0.21 & 0.26 
\enddata
\tablecomments{The mean offsets were calculated as average differences between the published index values and our
measurements for all observations of 9 calibration stars. 
The outliers (1 or 2 for each index) were excluded. 
Rms scatter about the mean is also given. 
The rms uncertainty per observation of the Lick calibrators 
(Worthey et al. 1994) is given in the last column.}
\end{deluxetable}


\begin{deluxetable}{lcccccc}
\tabletypesize{\scriptsize}
\tablecaption{CaT index.}
\tablewidth{0pt}
\tablehead{
\colhead{} &  
\colhead{CaT\,$(\mbox{\AA})$} 
}
\startdata
VUCD1        & 7.51 $\pm$ 0.32  \\
VUCD3/Strom547        & 6.92 $\pm$ 0.27 \\
VUCD4        & 5.71 $\pm$ 0.45  \\
VUCD5        & 7.97 $\pm$ 0.35  \\
VUCD6        & 6.13 $\pm$ 0.41  \\
VUCD7        & 6.37 $\pm$ 0.71  \\
Strom417    & 8.62 $\pm$ 0.45  \\
VCC1407     & 6.90 $\pm$ 0.51 \\
NGC4486B    & 7.32 $\pm$ 0.09  
\enddata
\end{deluxetable}

\clearpage


\end{document}